\documentclass[journal]{IEEEtran}
\usepackage[T1]{fontenc}
\usepackage[latin9]{inputenc}
\usepackage{cite}
\usepackage{array}
\usepackage{varioref}
\usepackage{float}
\usepackage{url}
\usepackage{multirow}
\usepackage{graphicx}
\usepackage[cmex10]{amsmath}
\usepackage{amsfonts}
\usepackage{bm}
\usepackage{xspace}

\makeatletter

\providecommand{\tabularnewline}{\\}
\floatstyle{ruled}
\newfloat{algorithm}{tbp}{loa}
\providecommand{\algorithmname}{Algorithm}
\floatname{algorithm}{\protect\algorithmname}

\newenvironment{lyxcode}
{\par\begin{list}{}{
\setlength{\rightmargin}{\leftmargin}
\setlength{\listparindent}{0pt}
\raggedright
\setlength{\itemsep}{0pt}
\setlength{\parsep}{0pt}
\normalfont\ttfamily}%
 \item[]}
{\end{list}}

\newcommand{\R}{\mathbb{R}}

\newcommand{\T}{\mathrm{T}}

\newcommand{\norm}{\mathcal{N}}

\newcommand{\xvect}{\mathbf{x}}
\newcommand{\yvect}{\mathbf{y}}

\newcommand{\lvect}{\mathbf{l}}

\newcommand{\mvect}{\mathbf{m}}
\newcommand{\zerovect}{\mathbf{0}}

\newcommand{\betavect}{\boldsymbol{\beta}}
\newcommand{\thetavect}{\boldsymbol{\theta}}

\usepackage{caption}
\usepackage{comment}
\usepackage{algorithmic}
\@ifundefined{showcaptionsetup}{}{%
 \PassOptionsToPackage{caption=false}{subfig}}
\usepackage{subfig}
\usepackage{color}
\makeatother

\newcommand{\revised}[1]{#1}
\newcommand{\rerevised}[1]{#1}
\newcommand{\rerevisedMaurizio}[1]{#1}

\newcommand{\ie}{i.e.\xspace}
\newcommand{\eg}{e.g.\xspace}
\newcommand{\etal}{et al.\xspace}
\newcommand{\adhoc}{ad hoc\xspace}
\newcommand{\etc}{etc.\xspace}

\begin{document}

\title{On User Availability Prediction\\and Network Applications}

\author{
  Matteo~Dell'Amico,
  Maurizio~Filippone,
  Pietro~Michiardi,
  and Yves~Roudier%
\thanks{M. Dell'Amico, P. Michiardi and Y. Roudier are with EURECOM, France.
E-mail: della@linux.it, pietro.michiardi@eurecom.fr, yves.roudier@eurecom.fr.}%
\thanks{M. Filippone is with the University of Glasgow, UK. E-mail: \mbox{maurizio.filippone@glasgow.ac.uk}.}%
\thanks{Manuscript received December 19, 2012; revised August 9, 2013 and February 17, 2014; accepted April 28, 2014.}
}

\maketitle

\begin{abstract}
User connectivity patterns in network applications are known to be
heterogeneous, and to follow periodic (daily and weekly) patterns.  In
many cases, the regularity and the correlation of those patterns is
problematic: for network applications, many connected users create
peaks of demand; in contrast, in peer-to-peer scenarios, having few
users online results in a scarcity of available resources. On the
other hand, since connectivity patterns exhibit a periodic behavior,
they are to some extent predictable. This work shows how this can be
exploited to anticipate future user connectivity and to have
applications proactively responding to it. We evaluate the probability
that any given user will be online at any given time, and assess the
prediction on six-month availability traces from three different
Internet applications. Building upon this, we show how our
probabilistic approach makes it easy to evaluate and optimize the
performance in a number of diverse network application models, and to
use them to optimize systems. In particular, we show how this approach
can be used in distributed hash tables, friend-to-friend storage, and
cache pre-loading for social networks, resulting in substantial gains
in data availability and system efficiency at negligible costs.
\end{abstract}

\begin{IEEEkeywords}
Predictive models, peer-to-peer computing, user availability
\end{IEEEkeywords}

\section{Introduction}

\revised{
\IEEEPARstart{I}{nternet} application workloads, being
direct consequence of human activity, are very often}
characterized by highly variable patterns of requests; daily, weekly
and seasonal patterns are ubiquitous in these applications, and have
been recorded in traces of file-sharing, instant messaging, and
distributed
computing \cite{bhagwan2003availability,chu2002availability,guha2006skype,javadi-et-al-setiathome-09,mickens2006exploiting,steiner2007kad}.
\revised{Despite this fact, in many cases system design and modeling
are performed without accounting for periodic request patterns that
are often very strong.}


The fact that user behavior is often correlated -- in the sense that
many users will behave similarly at the same time -- is especially
problematic.  {}``Flash crowds'' -- \ie, simultaneous requests
from many users at the same time -- create strains on application
resources; on the other hand, the fact that many users can go offline
at the same time creates problems in peer-to-peer applications, since
sudden and unexpected variations of available resources can
happen. Many applications are designed according to simple models of
user availability (\eg, modeling availability with a uniform
probability for all users \cite{bhagwan2003availability} or using
simple Markovian models \cite{duminuco2007proactive}) that fall short
when modeling the complexity of human behavior; in real world
scenarios, then, \revised{systems should be largely over-dimensioned
in order to avoid potential problems.}

\revised{
Over-dimensioned systems can re-use excess
resources in other ways. However, alternative usages (which need to be
exploitable at any moment in time) are likely to provide less added
value than those for which the system was designed in the first
place. In addition, even when elastic resource allocation is possible,
it is not necessarily instantaneous. For example, spawning and booting
new virtual machines can take some time -- in particular in cases of
heavy system load. Systems able to predict request spikes
can respond \emph{proactively} to increases in system
load -- allocating new resources when load is still tolerable --
rather than reactively.}

In this work, \rerevised{we model \emph{user
availability}, \ie whether users are online},
using
past statistics of \revised{their uptime} in order to capture idiosyncratic
behavior characteristics and obtain personalized predictions for each
of them. This allows us to design systems that respond proactively to
user uptime. We focus on characterizing future user availability with
an \emph{individualized},
\emph{long-term}, \emph{probabilistic}
\revised{
statistical model: a (scalable) system which assigns
a probability to the event of any user being online
at any time in the future.
}

\revised{
We evaluate our \rerevised{long-term predictions in different network}
applications. We \rerevisedMaurizio{use}} three different datasets
(Section \ref{sec:Datasets}) containing traces of user availability
spanning a time frame of about six months in the domains of instant
messaging, file sharing, and home gateways.

\revised{
Our statistical model (Section \ref{sec:Our-Model}) adopts logistic
regression to} combine several \revised{features} capturing different
periodic trends of individual and global user availability patterns.
By taking a probabilistic \rerevised{approach}, we are also able to
quantify the uncertainty in the estimation of model parameters and
account for it in predicting user availability.
\rerevised{This is often useful: for example, 
evaluating that a user will be online with probability 0.6 while for
another user the probability is 0.9 can allow applications to
\rerevisedMaurizio{make different decisions}
for the two users; conversely, a boolean predictor
that would just output ``online'' for both cannot allow \rerevisedMaurizio{for} such
differentiations.}

The scalability of our approach allows us to assess the performance of
our predictions on datasets of several weeks and with a large number
of users (Section \ref{sec:Prediction-Accuracy}).  The characteristics
of the dataset impact the quality of predictions that can be made, but
our model remains consistently useful in all cases; moreover, the
quality of predictions does not decrease substantially over time, even
after several weeks. \rerevised{The most relevant features are those
capturing individual and periodic long-term user behavior; this
suggests that our approach can be adopted whenever users have a
peculiar behavior which is stable over time (\eg, online social
networks, accesses to email servers, TV/radio consumption even over
IP, \etc).}

\revised{
\rerevised{Metrics for prediction quality, however, are not sufficient}
to understand
\rerevised{the impact that our method} could have on system
design. We show that in}
Section \ref{sec:Applications}, by presenting three use
cases in which our approach can be adopted to predict system
behavior and performance, and can be used to guide system choices.  By
driving node placement and data placement respectively in a
distributed hash table and a friend-to-friend storage application,
\revised{
resource usage can be reduced because
it is not necessary to perform excessive data replication to obtain
a desired level of data availability%
}; by adopting
connectivity predictions in social networks, the efficiency of cache
pre-fetching can be significantly improved.

\revised{
Within all classes of user behavior that can be predicted, in this
work we focus on user uptime because \textit{i)} traces are
availabile and \textit{ii)} as mentioned above, we can provide
concrete examples of exploiting availability prediction. In
Section~\ref{sec:Conclusions}, we conclude by mentioning other kinds of
user behavior and application use cases that could benefit from an
approach similar to the one described in this document.  }

\section{Related Work}

\label{sec:Related-Work}

Patterns of user availability are important in a large class of applications
since they impact the \emph{demand} of resources; they are essential
in peer-to-peer (P2P) systems, where they also determine the \emph{offer}
of available resources in the system. It is therefore unsurprising
that this issue has attracted particular interest in the P2P literature, as we discuss next.

\paragraph*{User Availability Modeling and Prediction}

Various papers
\cite{javadi-et-al-setiathome-09,kondo2008correlated,stutzbach2006churn}
focused on characterizing \emph{session length}, \emph{i.e.}  the
amount of time a user will spend online after connecting.
\rerevised{Predicting session length is useful} for cases where a
node's disconnection triggers expensive operations, such as data
maintenance in distributed hash
tables~\cite{rhea2004handling}\rerevised{; these techniques do not
  leverage on periodicity, but rather model session length with a
  probability distribution which is independent from the moment in
  which the session begins}. Our analysis is complementary to this,
since we focus on being able to predict connectivity patterns in the
long term.

Daily and weekly periodic behavior is a known feature of essentially
any trace of applications whose workloads depend on human factors.
This behavior has been reported, for example, in file-sharing
applications
\cite{bhagwan2003availability,chu2002availability,mickens2006exploiting,stutzbach2006churn},
instant messaging \cite{guha2006skype}, and distributed computing
\cite{javadi-et-al-setiathome-09}. In addition to merely recognizing
the presence of periodicity, in this work we exploit it in order to
enhance the quality of our predictions.

Mickens and Noble \cite{mickens2006exploiting} predict future node
uptime on several traces, including cases where availability patterns
are dictated mostly by technical reasons such as failures
(\eg, availability traces of nodes in PlanetLab). We
implemented this approach, and we devote Section~\ref{sec:mn} to show
how our approach outperforms it in terms of prediction accuracy,
scalability, and expressivity of the output.

\paragraph*{Applications}

Understanding availability is fundamental in P2P storage applications,
where data is uploaded redundantly
to various peers; when enough of them are online, data is
accessible by other peers. In several cases, availability is modeled
under the assumption that the probability that any node is online at
any time is constant. In order to avoid problems due to correlated
downtimes, this probability has to be estimated very conservatively,
resulting in a dramatic system overprovisioning (\eg, Bhagwan
\etal~\cite{bhagwan-et-al-04} adopt the lowest observed
fraction of connected users in the history of the
system). \rerevised{In some
  approaches~\cite{mickens2006exploiting,pamies2010towards},
  heterogeneity in availability traces is exploited to} increase load
on nodes with higher availabilities \rerevised{and} optimize system
performance. These approaches inevitably result in an imbalance on
requested resources, penalizing the most available nodes. Unlike them,
we show that better performance can be obtained without requiring
additional resources from any node. Pamies-Juarez \etal~\cite{pamiesjuarez11enforcing} show how better system performance can
be obtained without requiring usage of more resources from more
available nodes, but they validate their findings with synthetic
experiments that do not account for correlation in uptime, which is
extremely problematic for the envisioned storage application. Finally,
Kermarrec \etal~\cite{KERMARREC:2010:HAL-00521034:2} propose
\revised{an} \adhoc method for data placement accounting for
periodic availability patterns. In Section \ref{sec:Applications}, we
show two use cases where optimization strategies driven by
availability predictions applied to peers can substantially optimize
data availability; in our scenario, our method largely outperforms the
one \revised{by Kermarrec \etal}

User request patterns have regularities that can be used to anticipate
future user behavior and plan capacity accordingly. G\"ursun
\etal~\cite{gursun2011describing} use a simple model of these
regularities to predict the total number of users that will request an
individual video file on YouTube \textit{the next day}. With respect
to their work, we perform our predictions on a personalized basis, and
several weeks in advance.
In Section~\ref{sub:sn-cache-prefetching} we show how our predictions
can be used to drive a pre-fetching strategy for newsfeeds in social
networks: \revised{in this case per-user predictions are needed, and
non-personalized predictive
approaches~\cite{lempel2003predictive,wu2002prediction} cannot be
applied.}


\revised{Regularities of user requests are not limited to temporal
  patterns: user location and social connectivity can also be used to
  drive smart caching and pre-fetching
  strategies~\cite{scellato2011track,traverso2012tailgate}. While
  considering these additional features is not within the scope of
  this work, these efforts confirm that proactive approaches to system
  design that anticipate user behavior are feasible and efficient. We
  think that a probabilistic approach similar to the one we present
  here could also be applied in the design of systems such as those
  ones.}

\section{Datasets}

\label{sec:Datasets}

We evaluate our predictions on three availability traces (\ie,
informations about when users are online) from Internet applications.
These traces share three key properties that make
them suitable for our study.

First, they comprise \revised{thousands of} users and they are long
enough (at least six months each) to let us test the quality of
predictions over several weeks.

Second, \revised{uptime depends on which motivations users have to
  connect: the three datasets we use allow us to examine cases in
  which the different nature of the application results in different
  motivations for users to be online. \rerevised{In addition to the
    simple fact that users have to be connected to use the service,
    in the Kad dataset there is an incentive mechanism that encourages users to
    stay online even more.}}


Third, the traces we use \rerevised{present} low or no sampling
bias~\revised{\cite{stutzbach2009unbiased,gjoka2010walking}}, since they are
essentially uniform or full samples of the users in each system.

To obtain a uniform duration between datasets, we consider only the
first 24 weeks. We use the first 18 weeks to \revised{extract our features},
and the remaining 6 weeks to test \revised{the predictive performance of our model}: more details
can be found in Section \ref{sec:Our-Model}. We expressed each trace
as an availability matrix $A$ by sampling the availability of each
user with a period of one hour. We denote $A_{u,t}=1$ if user $u$
is available at timeslot $t$, and $A_{u,t}=0$ otherwise.

The GW and Kad datasets are publicly available; since IM may contain
potentially sensitive informations about user behavior, we cannot
commit on making it public. To help reproducibility and re-use of our
experiments, our source code is available in a public GIT
repository.\footnote{\url{https://bitbucket.org/matteodellamico/uptime-prediction}}

\subsection{IM -- Instant Messaging}

The \emph{IM} trace is extracted from the server logs of an instant
messaging server in Italy; one of the authors is an administrator of
the server.  The trace contains the complete log of
connection/disconnection events on the server in the period between
January 10, 2010 and June 27, 2010; $1,174$ distinct users connected
to the server in that time-frame.  We denote a user as {}``online'' if
at least one client software is logged in with the user's credentials,
regardless of the status set in the client (\emph{e.g.}, available,
busy, invisible\ldots{}).

In this system, very strong daily patterns are observable. We
attribute this to two reasons: first, most users live in the same
timezone; second, \rerevised{they are likely to go online whenever
  they are in front of their computer, in order to be reachable.}

In the period of May 18--20, \rerevised{a large number of accounts was
  registered by automatic tools; such accounts, being non-human, had
  connectivity patterns that were fundamentally different from those
  observed in the rest of the trace. Separating human and non-human
  behavior with means such as CAPTCHAs and characterizing online
  behavior of automated tools are both outside the scope of this
  work. This phenomenon, however, does not affect the results shown in
  this work:
  these accounts were registered only after our training period (\ie,
  in the last six weeks of the trace).} Availability predictions for
them have therefore neither been generated nor tested.


\subsection{GW -- Gateways}
\label{sec:gw}

The \emph{GW} trace has been extracted by Serge Defrance \emph{et
al.}~\cite{defranceefficient}.
It comprises traces from $24,781$ residential gateways of the French
ISP \emph{Free} having a fixed IP address, and it was
obtained by pinging each gateway every 10 minutes. The set of IP
addresses was chosen randomly within the address space allocated for
these gateways, and a gateway was included in the trace if it had
answered to the ping message at least once. In this work we consider
availabilities in the 24-week interval between June 29, 2010 and
December 24, 2010\rerevised{; four aberrations where most users appear
  offline due to measurement artifacts are present in the dataset and
  are acknowledged by the
  authors}.\rerevised{\footnote{\url{http://www.thlab.net/~lemerrere/trace_gateways/}}}

The gateways provided by Free offer access to telephone services, TV
and DVR in addition to Internet access; \rerevised{for this reason
  many users keep their gateways almost always on. As a result,
  the average percentage of connected users is 86\% over the trace.}
Daily and weekly periodic behavior is still \rerevised{present}, with a number of
users that disconnect their gateways mostly during the night.

\subsection{Kad}

The \emph{Kad} trace has been extracted by Steiner \emph{et al.}
\cite{steiner2007kad}, measuring the connectivity on nodes on the Kad
distributed hash table, which is used by clients of the popular
eDonkey2000 network. In Kad, nodes are identified by a randomly
generated 128-bit identifier called Kad ID; the trace contains the
result of a \emph{zone crawl} resulting in the availability trace
(sampled every 5 minutes) of all the $400,375$ nodes sharing a common
8-bit prefix appearing online in a six-month period. We consider
availability data between September 23, 2006 and March 10, 2007.

In the eDonkey2000 network, users are implicitly motivated to stay
online when they are downloading files; moreover, there is an elaborate
\rerevised{credit-based} incentive scheme~\cite{caviglione2008emule}
designed to reward users that stay online and upload data to peers
with faster download speeds.

\subsection{First Observations}

\begin{figure}
\begin{centering}
\includegraphics[width=\columnwidth]{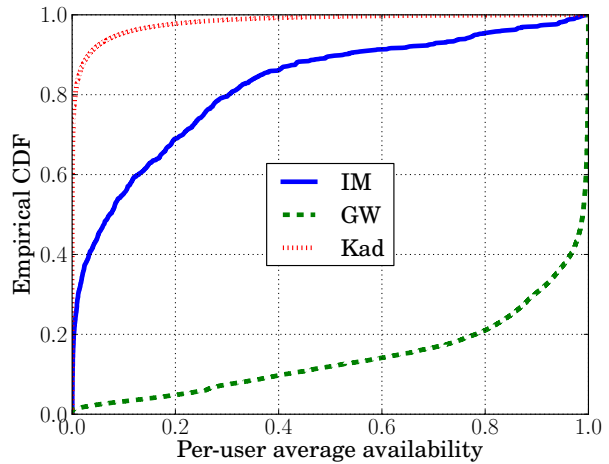}
\par\end{centering}

\caption{CDF of per-user average availability.}
\label{fig:CDF-of-per-node}
\end{figure}

\begin{figure*}
\begin{centering}
\subfloat[IM]{
\includegraphics[width=0.32\textwidth]{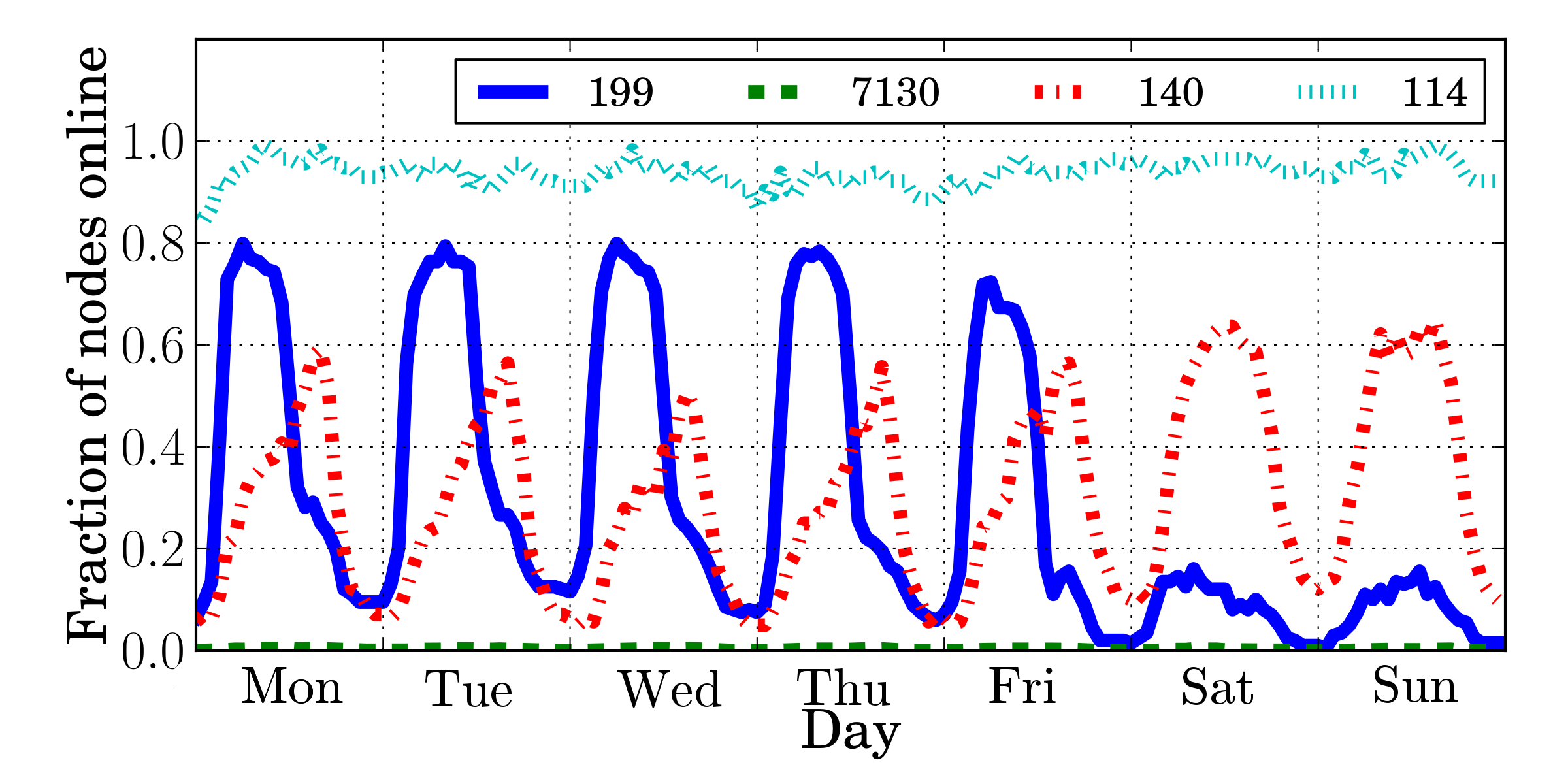}
}
\subfloat[GW]{
\includegraphics[width=0.32\textwidth]{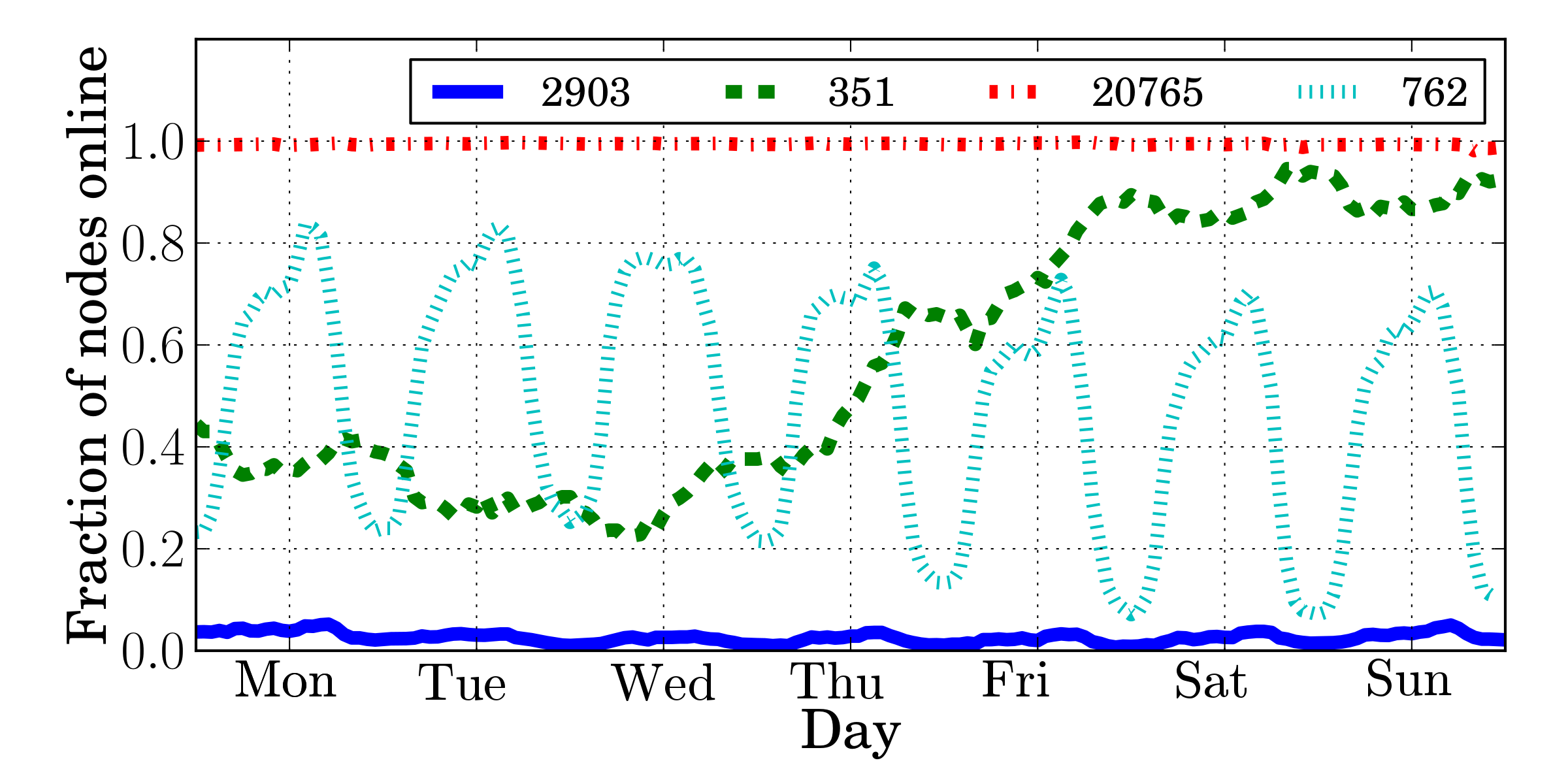}
}
\subfloat[Kad]{
\includegraphics[width=0.32\textwidth]{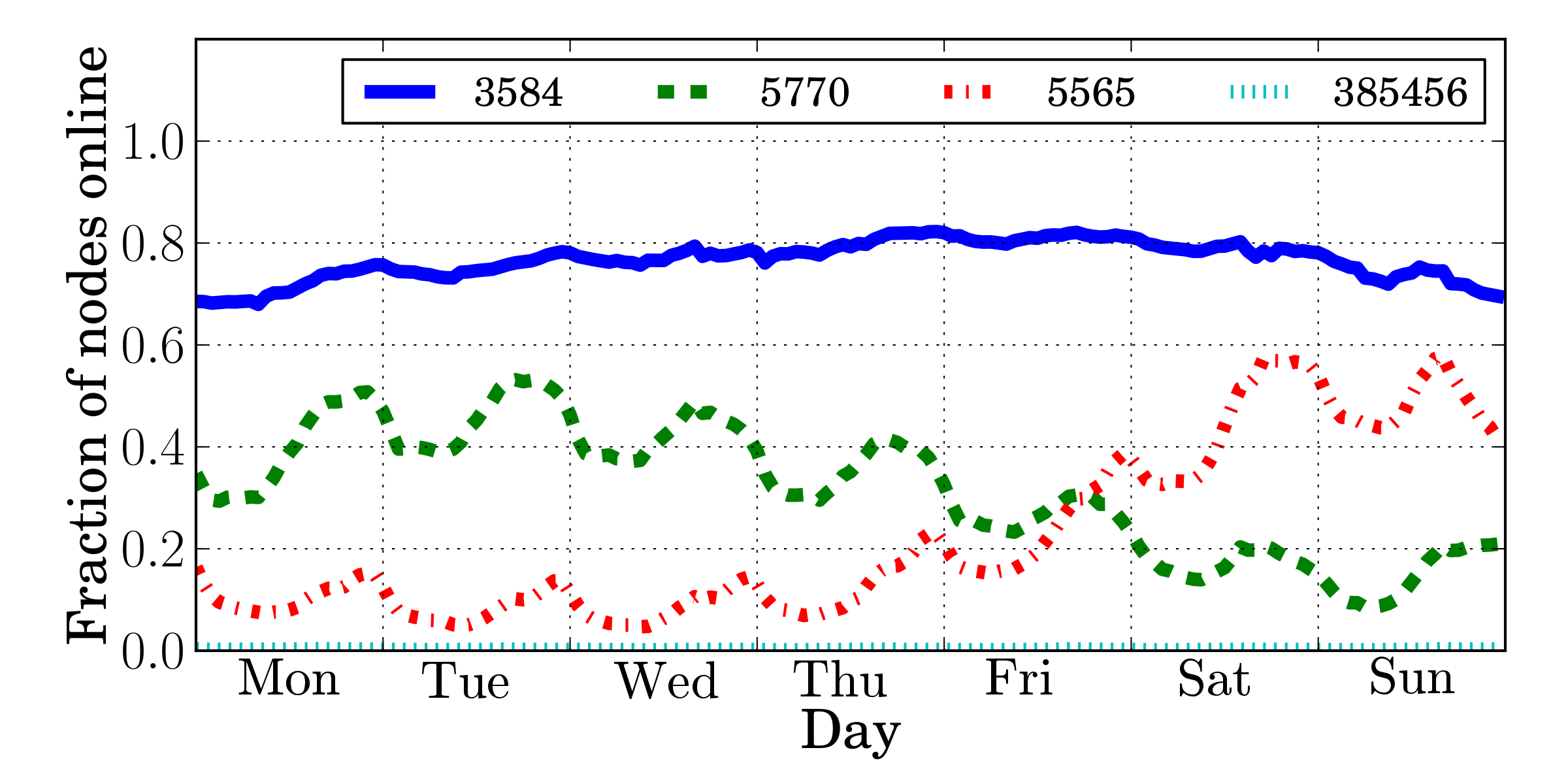}
}
\caption{\rerevised{Availability patterns clustered through $k$-means. One week of data shown per dataset.}}
\label{fig:One-typical-week}
\end{centering}
\end{figure*}






In Figure~\vref{fig:CDF-of-per-node}, we show the distribution of the
per-user average availability in each dataset (\ie, the distribution
of the fraction of time each user spends online). Because of the
applications' nature, these values are drastically different.
\rerevised{It is interesting to note that, even though an incentive
  scheme has been devised in order to convince users to stay online
  more often, the Kad trace is the one with lowest availability values
  overall. It appears that the incentive mechanism present in the
  network has a limited impact, and that the very nature of the
  applications that encourages users to stay online to remain
  reachable to receive instant messages (IM) or phone calls (Kad)
  plays a much stronger role.}

\revised{In Figure~\vref{fig:One-typical-week}, we highlight the
  diversity of per-user availability in each \rerevised{dataset}. We
  took a sample week from each trace, and sampled the availability of
  each user with a granularity of one hour\rerevised{, to obtain a
    matrix where lines represents users and columns timeslots. Each
    cell indicates whether the user was
    online or not at that time.}

 We performed $k$-means clustering on \rerevised{the lines of this
   matrix, resulting in a clustering of users} (\rerevised{we chose
   $k=4$ for plot readability}), and we plot the average
 number of per-cluster available users. Each line corresponds to a
 cluster; in the legend, we report the number of users belonging to
 each cluster.

\revised{From Figure~\ref{fig:One-typical-week}, we can obtain an
  intuitive grasp of the behavior of some representatives groups of
  users in the cluster.}  In all cases there are clusters of users who
are mostly on and mostly off, and others with different inclinations
to connect: mostly during weekdays or mostly during
weekends. \rerevised{The size of these clusters (shown in the legend)
  can explain the distribution of per-user average availability
  values shown in Figure~\vref{fig:CDF-of-per-node}.}

In the IM trace, it is interesting to note that some users tend to
connect during office hours \rerevised{(blue cluster)}, and others in
free time (early night and weekends\rerevised{, red cluster}). \rerevised{The
clustering results show that such a phenomenon is less common for GW
and Kad, where the $k$-means clustering instead captured groups of
users whose availability pattern changed during the week under scrutiny (green cluster for GW, red and green clusters for Kad).}}


\rerevised{In summary, our datasets represent three different types of
  network applications with markedly different user behavior;
  moreover, IM clients~\cite{bonfiglio2009detailed}, home
  gateways~\cite{nada,defranceefficient}, and file-sharing
  clients~\cite{steiner2007kad} are all popular platforms for running
  peer-to-peer applications, making these datasets ideal to test the
  peer-to-peer application use cases shown in
  Sections~\ref{sub:app-dht} and~\ref{sub:app-storage}.}

\section{Our Model}

\label{sec:Our-Model}

In this section we introduce the model that we use for predicting
user availability. We start with a set of \revised{features} that
identify global and personalized periodic trends, and combine them in
a classifier
, as we will see shortly.  In this work we adopt 
a \rerevisedMaurizio{fully} probabilistic classifier based on logistic regression, as it
provides high descriptive power and is quite flexible.  We split the
data sets in four non-overlapping consecutive periods A, B, C, and D,
each of duration of six weeks.  Inference on the parameters of the
classifier is carried out using A to compute the features and B to
obtain the corresponding labels.  We use C and D to assess the
predictive performance of the probabilistic classifier.  Now C is used
to construct \revised{features} and D to obtain
labels against which classification performance is assessed.

The proposed procedure avoids any issues with reusing the same data
for inference and for assessing the quality of the classifier, and
preserves temporal information in the process as if we were to apply
this procedure in a real scenario.  Although this does not allow to
evaluate some sort of cross-validation error, we report results on
three different scenarios comprising \revised{thousands of} users,
that substantiate the hypothesis that patterns of availability
are \revised{largely} predictable.

In several cases stemming from P2P
applications\rerevised{~\cite{beverly2003designing,bonfiglio2009detailed}},
it is most interesting to consider ``superpeers'', which are nodes
with a relatively high availability: for example, in the Wua.la
distributed storage application~\cite{wuala07}, data is stored on
peers that have spent at least an average of 4 hours per day online in
the past week.  Mirroring this latter requirement, we created
restricted \emph{filtered} datasets containing nodes that spend on
average at least 4 hours per day online. This filtering is performed
on period A for both period A and B, and on period C for periods C and
D: in other words, a node will appear in the filtered dataset in
period B (resp. D) if it has been available on average at least four
hours per day in period A (resp. C). \rerevised{This process resulted
in selecting 405 users in the IM dataset for the A-B periods, and 408
nodes in the C-D periods. For GW, the selected nodes are 22,620 (A-B
periods) and 23,184 (C-D). For Kad, they are 11,472 (A-B) and 12,522
(C-D).}

Filtered datasets allow
us to evaluate the predictor quality for use cases of interest
ignoring easily predictable, mostly-offline users. We use the
filtered datasets for our application use cases in Sections
\ref{sub:app-dht} and \ref{sub:app-storage}.


\subsection{\revised{Features}}
\label{sec:features}

User availability follows periodic daily and weekly trends \revised{which}
affect the whole system. Moreover, each user has his/her own
particular daily and weekly trends (\emph{e.g.} being more or less
likely to connect on nights and on weekends\revised{, as shown in
Figure \vref{fig:One-typical-week}}), as well as different likelihood
to be online at all (see Figure \vref{fig:CDF-of-per-node}). We
designed a set of
\revised{features} aimed to capture each of these trends.

We are interested in creating a prediction for each user and each
timeslot in the future. For a user $u$ and a timeslot $t$, we identify
a set of observations that are related to the trend we want to
identify with the \revised{feature} at hand, and we simply count the numbers
$n_{\mathrm{on}}$ and $n_{\mathrm{off}}$ of observations in which we find nodes
respectively online and offline. We then return as an output of $i$-th
\revised{feature} the value
$$
f^{(i)}_{u,t} = \frac{n_{\mathrm{on}}+1}{n_{\mathrm{on}}+n_{\mathrm{off}}+2}.
$$
\revised{These values} can be seen as the expected value of the posterior of a
Bernoulli trial assuming a flat prior
$\mathrm{Beta}(1,1)$. \revised{We point out that it is computationally
trivial to update these features when new observations become available.}

We define five different \revised{features}, which
differ \rerevised{by set of users considered (individual or global)
and by periodicity (daily, weekly, and flat). Features differ} only by
the definition of the set of observations which is taken into account
in order to compute $f^{(i)}_{u,t}$:
\begin{enumerate}
\item \emph{Global daily}. Observations of all users, same
  time of day of $t$ (\emph{e.g.}, if $t$ is at 9:00 AM,
  observations taken into account will be at all days at 9:00 AM).
\item \emph{Global weekly}. All users, same time of day and same day of
  week of $t$ (\emph{e.g.}, if $t$ is on a Monday at 9:00 AM,
  observations taken into account will be at all Mondays at 9:00 AM).
\item \emph{Individual flat}. Exclusively user $u$, any observation in
  the training set.
\item \emph{Individual daily}. User $u$, same time of day of $t$.
\item \emph{Individual weekly}. User $u$, same time of day and same day of
  week of $t$.
\end{enumerate}

\revised{It is possible to consider additional features accounting for
periodic behavior. We observed a \rerevised{sharp} decrease in the number of
connections in holiday periods (\ie, the month of August and the
Christmas period). Unfortunately, our traces are not long enough to
allow us taking into account seasonal variations. We consider it more
than reasonable to assume, however, that even longer traces would
result in increased accuracy. 

In addition, we empirically observed that public holidays result in
availability patterns that are similar to those of
weekends. Unfortunately, since our traces comprise users from
different countries, it is difficult to define precisely which days
are public holidays.}

\subsection{Logistic Regression}

The five \revised{features} above are able to capture different
aspects of patterns of availability, and we are interested in
combining these pieces of information to accurately predict future
user availability.  Moreover, we aim to assess the relative
importance of the five \revised{features} in doing so.  We propose a
\rerevisedMaurizio{fully} probabilistic 
logistic regression classifier~\cite{Bishop06}.
\rerevisedMaurizio{Adopting a probabilistic classifier yields a probability of users to be online at a given time.
This reflects in the possibility to obtain a degree of how certain the classifier is on the availability of users that can be exploited in network applications.
}

In logistic regression, and more in general in a classification
problem, a set of labels $\yvect = \{ y_1, \ldots, y_n\}$ is
associated with a set of samples $\{ \xvect_1, \ldots, \xvect_n\}$,
each described by a set of $d$ \rerevisedMaurizio{features} so that $\xvect_i \in \R^d$.
In our application the labels $y_i \in \{ 0, 1\}$ represent $A_{u,t}$,
that is the availability of user $u$ at time $t$, and $\xvect_i$ is
the corresponding set of five \revised{features} $\left(f^{(1)}_{u,t}, \ldots,
f^{(5)}_{u,t}\right)$.  The logistic regression classifier models the labels
$y_i$ as conditionally independent, and as draws from a Bernoulli
distribution with probability of ``success'' given by: 
$$
p(y_i = 1 | \xvect_i) = \sigma\left( \beta_0 + \sum_{j=1}^d (\xvect_i)_j \beta_j\right),
$$
where $\sigma$ denotes the logistic function $\sigma(a) = \frac{1}{1 + \exp(-a)}$.
Note that we have introduced \revised{an intercept} term $\beta_0$ in the linear combination to allow for a term in the combination independent from the \rerevisedMaurizio{features}.
In order to keep the notation uncluttered, we define the set of weights of the combination together with the \revised{intercept} term as $\betavect = (\beta_0, \beta_1, \ldots, \beta_d)$, and with abuse of notation we redefine $\xvect_i = (1, \xvect_i)$ so that the linear combination becomes $\xvect_i^{\T} \betavect$.
Also, we define $l_i^{+} := p(y_i = 1 | \xvect_i) = \sigma( \xvect_i^{\T} \betavect)$ and similarly $l_i^{-} := p(y_i = 0 | \xvect_i)$.
Finally, let $X$ be the $n \times (d+1)$ matrix obtained by stacking the vectors $\xvect_i$ by row.

\rerevisedMaurizio{The goal of a classification approach is to tune or infer the parameters $\betavect$ based on past user availability data, and use this to predict future user availability.
It is worth noting here that, once the parameters $\betavect$ are inferred from data, they offer an interpretation of the relative importance of the different \rerevisedMaurizio{features} in discriminating between the classes.
We now report the approach that we take to infer the parameters $\betavect$ and explain how we predict future user availability. 
}

\subsubsection{Bayesian Inference}

\rerevisedMaurizio{
In this work we take a Bayesian approach to infer the parameters of the logistic regression classifier.
The motivation for doing so is that the Bayesian approach offers a sound quantification of uncertainty in parameter estimates and in predictions, as demonstrated in several applications~\cite{FilipponeAOAS12}.
}
In a probabilistic setting, the predictive distribution for a new test sample $\xvect_*$ is obtained via the following marginalization:
\begin{equation} \label{eq:logreg:predictive}
p(y_* | \yvect, X, \xvect_*) = \int p(y_* | \xvect_*, \betavect) p(\betavect | \yvect, X) d\betavect.
\end{equation}
Equation \ref{eq:logreg:predictive} requires the so called posterior distribution over $\betavect$ after observing $X$ and $\yvect$.
Note that this expression is quite appealing as the uncertainty on the parameters a posteriori is effectively averaged out when predicting future availability.
As a result, predictions no longer depend on model parameters.
This is in contrast to what would happen if we optimized the parameters, obtaining say $\hat{\betavect}$, that would lead to predictions of the form $p(y_* | \yvect, X, \xvect_*, \hat{\betavect})$ that are conditioned on a specific choice of $\betavect$.
\rerevisedMaurizio{The core of quantification of uncertainty in parameter estimates and predictions is therefore in the posterior distribution over $\betavect$.
By Bayes theorem, $p(\betavect | X, \yvect)$ is proportional to $p(\yvect | X, \betavect) p(\betavect)$, namely the product of 
the likelihood of the observed availability given the parameters, and the prior distribution over the parameters.
}
The log-likelihood of the observations given the parameters is
$$
\log[p(\yvect | \betavect, X)] = \sum_{i=1}^n \left[ y_i \log(l_i^{+}) + (1 - y_i) \log(l_i^{-}) \right].
$$
We assign a Gaussian prior over $\betavect$ with mean $\mvect_0$ and \rerevisedMaurizio{features} $S_0$.
In this work we assume a rather flat prior over $\betavect$; in particular, we set $\mvect_0 = \zerovect$ and $S_0 = \alpha I$ with $\alpha = 10^4$.
Given the amount of data and the variance of the prior, the Gaussian assumption is not restrictive, but has the advantage of being simple to deal with.

Given the form taken by the likelihood in the logistic regression model, it is not possible to obtain the posterior distribution on $\betavect$ in closed form, and it is therefore necessary to resort to some approximation.
It is possible to approximate the posterior distribution using deterministic approximations~\cite{Bishop06,FilipponeCSDA11}, or employ a Markov chain Monte Carlo (MCMC) approach to obtain samples from the posterior distribution~\cite{Robert05}.
\rerevisedMaurizio{Here we employ a deterministic Gaussian approximation technique called Laplace Approximation.}

\subsubsection{Laplace Approximation}

The LA algorithm is a deterministic technique that approximates a function using a Gaussian.
This approximation places the approximating Gaussian at the mode of the target distribution and sets its covariance to the covariance of the target distribution~\cite{Bishop06}.
Although it can be argued that a Gaussian approximation can be very poor when the target distribution is far from being Gaussian, in the limit of infinite data the posterior distribution will tend to a Gaussian (see, e.g., \cite{Kass89} for further details).
In this application we are dealing with very large amounts of data (in the order of millions) and therefore the LA is appropriate in characterizing the uncertainty on parameter estimates.
Moreover, given that the complexity of the whole algorithm is linear in the number of data (see below), it scales well for large data sets.

In this work, we make use of the \rerevisedMaurizio{iterative} Newton-Raphson optimization strategy \rerevisedMaurizio{to carry out the Gaussian approximation}, as it is commonly employed in logistic regression.
The Newton-Raphson optimization \rerevisedMaurizio{iteratively} uses gradient and Hessian of the \revised{logarithm of the target function to locate its mode}. 
Denoting with $\betavect^{\prime}$ the update of the parameters after one iteration, the Newton-Raphson formula is
$$
\betavect^{\prime} = \betavect - (\nabla_{\betavect} \nabla_{\betavect} \mathcal{L})^{-1} \nabla_{\betavect} \mathcal{L}.
$$
This update is repeated until the gradient vanishes, which in practice is assessed by checking that the norm of the gradient is less than a \revised{given} threshold.
\rerevisedMaurizio{In the case of logistic regression, the update equation is:} 
$$
\betavect^{\prime} = \betavect + (X^{\T} \Lambda X + S_0^{-1})^{-1} [X^{\T} (\yvect - \lvect^{+}) - S_0^{-1} (\betavect - \mvect_0)],
$$
\rerevisedMaurizio{where we have defined $\Lambda$ as a diagonal matrix with $\Lambda_{ii} = l_i^{+} l_i^{-}$.}
After convergence, we have an approximation $p(\betavect | \yvect, X) \simeq \norm(\betavect | \mvect, S)$, where $\mvect$ is the value of $\betavect$ at the end of the optimization and $S = (X^{\T} \Lambda X - S_0^{-1})^{-1}$ is the \rerevisedMaurizio{covariance of $\mathcal{L}$} evaluated at $\mvect$.
The complexity of each update of $\betavect$ is linear in the number of samples $n$, and cubic in the number of parameters due to the inversion of a $(d+1)\times(d+1)$ matrix. 
Given that in our application $d=5$, the LA results in a very fast method (reported in Algorithm \vref{alg:logreg:inference}) for inferring the parameters of the model.

Note that the probabilistic framework allows one to carry out the inference by processing data in batches.
In particular, this can be accomplished by simply treat the posterior on $\betavect$ after processing one batch of data as the prior for $\betavect$ before processing a new batch of data.
We resorted to this option in the case of the KAD data set only, where the data were split in ten batches of several tens of millions of samples due to limitations in storing all the data \revised{into} memory.

\begin{algorithm} \caption{LA for inference in Logistic Regression}\label{alg:logreg:inference} 
\textbf{Input}: Data: $X$, $\yvect$ - Prior: $p(\betavect) = \norm(\betavect | \mvect_0, S_0)$ - stopping criterion $\thetavect$

\textbf{Output}: Gaussian posterior: $p(\betavect | \yvect, X) \simeq \norm(\betavect | \mvect, S)$

~

$\betavect = \zerovect$

\textbf{while} $\left( \| \nabla_{\betavect} \mathcal{L}(\betavect) \| > \theta \right)$ 

\hspace{0.5cm} $\betavect^{\prime} = \betavect + (X^{\T} \Lambda X + S_0^{-1})^{-1} [X^{\T} (\yvect - \lvect^{+}) - S_0^{-1} (\betavect - \mvect_0)]$

\hspace{0.5cm} $\betavect = \betavect^{\prime}$

\textbf{return} $\mvect = \betavect, S = \nabla_{\betavect} \nabla_{\betavect} \mathcal{L}(\mvect)$
\end{algorithm}














\subsubsection{Predictions}

The predictive distribution is obtained by solving the integral in Eq. \ref{eq:logreg:predictive}, where now we have a Gaussian approximation to $p(\betavect | X, \yvect) \simeq \norm(\betavect | \mvect, S)$.
Defining $m_a = \xvect_*^{\T} \mvect$ and $s^2_a = \xvect_*^{\T} S \xvect_*$, the predictive distribution results \rerevisedMaurizio{can be approximated} by 
$$
p(y_* = 1 | \yvect, X, \xvect_*) \simeq \sigma\left[ (1 + \pi s_a^2 / 8)^{-1/2} m_a\right].
$$
The complexity of evaluating this expression (in Algorithm \vref{alg:logreg:predictions}), once $\mvect$ and $S$ are available, is linear in the number of parameters.

\begin{algorithm}  \caption{Predictions with Gaussian posterior}\label{alg:logreg:predictions} 
\textbf{Input}: Data: $X$, $\yvect$ - Gaussian posterior: $p(\betavect | \yvect, X) \simeq \norm(\betavect | \mvect, S)$ - test data: $\xvect_*$

\textbf{Output}: Prediction $p(y_* = 1 | \yvect, X, \xvect_*)$

~

$m_a = \xvect_*^{\T} \mvect$

$s^2_a = \xvect_*^{\T} S \xvect_*$

\textbf{return} $p(y_* = 1 | \yvect, X, \xvect_*) \simeq \sigma\left[ (1 + \pi s_a^2 / 8)^{-1/2} m_a\right]$
\end{algorithm}

\rerevised{
\subsection{Scalability}

To compute predictions, our method performs two steps: \emph{feature
extraction} and \emph{logistic regression}. Computation is
lightweight: both steps scale linearly and they are dominated by the
speed at which input data is read and parsed. The feature extraction
component, which is the one dealing with the largest input, can be
easily parallelized; the logistic regression component, in our
prototype implementation, can handle the around 100 million data
points of the GW dataset in around one minute.}









\section{Prediction Accuracy}

\label{sec:Prediction-Accuracy}


We report two measures of performance for our classifier, namely the
Area Under the Curve (AUC) of the Receiver Operating Characteristic
(ROC) and the geometric mean of the likelihood on test data (GM).

\rerevised{AUC measures how well a probabilistic classifier balances
  true and false positives, with the optimal value of 1 corresponding}
to the \rerevised{case} where all true positives are associated
probabilities larger than \rerevised{any} true \rerevised{negative}.

The GM score, instead, is calculated as
\begin{eqnarray}
\mathrm{GM} & = & \sqrt[|U||T|]{\prod_{u \in U,t \in T} l(u,t)} \notag \\
& = &\exp\left(\frac{1}{|U||T|}\sum_{u \in U,t \in T}\log l(u,t) \right), \notag
\end{eqnarray}
where $l(u,t)$ is the likelihood of the availability observation
$A_{u,t}$ for user $u$ and time $t$ given a predicted probability of
$P_{u,t}$: $l(u,t)=P_{u,t}$ if $A_{u,t}=1$, else $l(u,t)=1-P_{u,t}$.
GM is a value between $0$ and $1$ and it is larger for classifiers
that assign correct predictions with high confidence, measured through
the predictive probability.  GM complements the AUC metric by
accounting for the confidence level of the output of the classifier.

\begin{table*}
\centering
\begin{tabular}{l|rrr|rrr|rrr}
& \multicolumn{3}{c|}{IM} & \multicolumn{3}{c|}{GW} & \multicolumn{3}{c}{Kad} \\
& AUC & GM & $\betavect$ (st. dev.) & AUC & GM & $\betavect$ (st. dev.) & AUC & GM & $\betavect$ (st. dev.) \\
\hline
\textbf{ALL} & \textbf{.939} & \textbf{.785} & & \textbf{.916} & \textbf{.807} & & \textbf{.826} & \textbf{.933} & \\
Individual daily & .938 & .779 & 0.47 (.01) & .912 & .806 & 1.04 (.00) & .826 & .933 & 0.38 (.00) \\
Individual flat & .914 & .756 & 0.52 (.01) & .915 & .805 & 0.17 (.00) & .816 & .931 & 0.12 (.00) \\
Individual weekly & .918 & .776 & 0.60 (.01) & .896 & .803 & 0.04 (.00) & .755 & .931 & 0.11 (.00) \\
Global daily & .583 & .617 & -0.06 (.01) & .522 & .659 & 0.02 (.00) & .532 & .914 & -0.06 (.01) \\
Global weekly & .593 & .618 & 0.29 (.01) & .523 & .659 & -0.01 (.00) & .530 & .914 & 0.14 (.01) \\
\multicolumn{10}{c}{}\\
& \multicolumn{3}{c|}{IM filtered} & \multicolumn{3}{c|}{GW filtered} & \multicolumn{3}{c}{Kad filtered} \\
& AUC & GM & $\betavect$ (st. dev.) & AUC & GM & $\betavect$ (st. dev.) & AUC & GM & $\betavect$ (st. dev.) \\
\hline
\bf ALL & \bf .901 & \bf .666 & & \bf .845 & \bf .823 & & \bf .730 & \bf .599 &  \\
Individual daily & .890 & .652 & 0.54 (.01) & .834 & .822 & 0.71 (.00) & .728 & .599 & 0.65 (.00) \\
Individual flat & .815 & .611 & 0.28 (.01) & .841 & .821 & 0.18 (.00) & .691 & .589 & 0.05 (.00) \\
Individual weekly & .889 & .653 & 0.69 (.01) & .802 & .819 & 0.02 (.00) & .706 & .592 & 0.20 (.00) \\
Global daily & .612 & .510 & -0.17 (.01) & .539 & .762 & 0.02 (.00) & .532 & .560 & -0.21 (.00) \\
Global weekly & .627 & .513 & 0.32 (.01) & .543 & .760 & -0.01 (.00) & .534 & .560 & 0.23 (.00)
\end{tabular}

\caption{Prediction accuracy.\label{tab:accuracy}}
\end{table*}

In Table~\vref{tab:accuracy}, we show our error metrics evaluated on
all datasets, for our combination of all \revised{features}, along
with a breakdown -- for each of them -- of the error metrics
achievable by using a single \revised{feature} in the \revised{classifier}. 
We also report the mean and the standard deviation of the
posterior over $\betavect$ obtained by the logistic regression
classifier using all the \revised{features} normalized to have unit
variance.

From Table~\ref{tab:accuracy}, we notice that the combined predictor
is consistently better or equivalent to the other ones, showing that
the combination effectively integrates information from the five
\revised{features} that capture basic trends.  Error metrics are worse on
datasets like IM filtered and Kad filtered, because they lack nodes
that are mostly offline and that are therefore easier to predict.
This effect is stronger on the GM metric, showing that the classifier
correctly assigns very low probabilities of connectivity to nodes that
are mostly offline, and it is less confident on nodes with more
irregular behavior.

Table~\ref{tab:accuracy} also shows that global \revised{features}
consistently perform worse than individual ones: even if global trends
can be clearly identified and they are useful in characterizing
availability, it is clear that information about individual trends is
simply more important.  \rerevisedMaurizio{However, \rerevised{the
    accuracy of the classifier} using all the features is higher than
  the one for the classifiers using a single feature\rerevised{,
    indicating} that combining \rerevised{all} the features using logistic
  regression is beneficial.}

\begin{figure}
\begin{center}
\begin{tabular}{ccc}
\includegraphics[width=0.46\columnwidth]{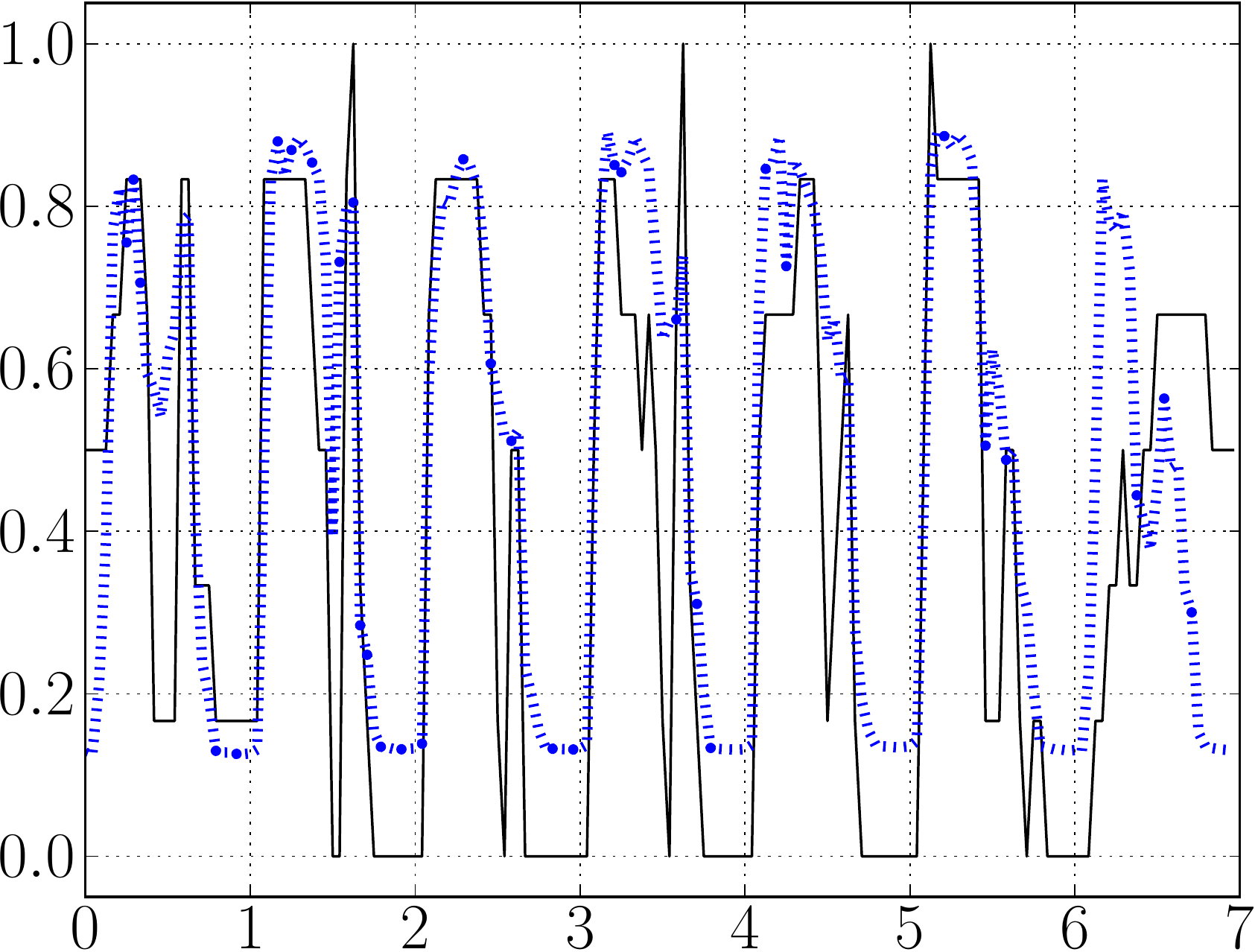} &
\includegraphics[width=0.46\columnwidth]{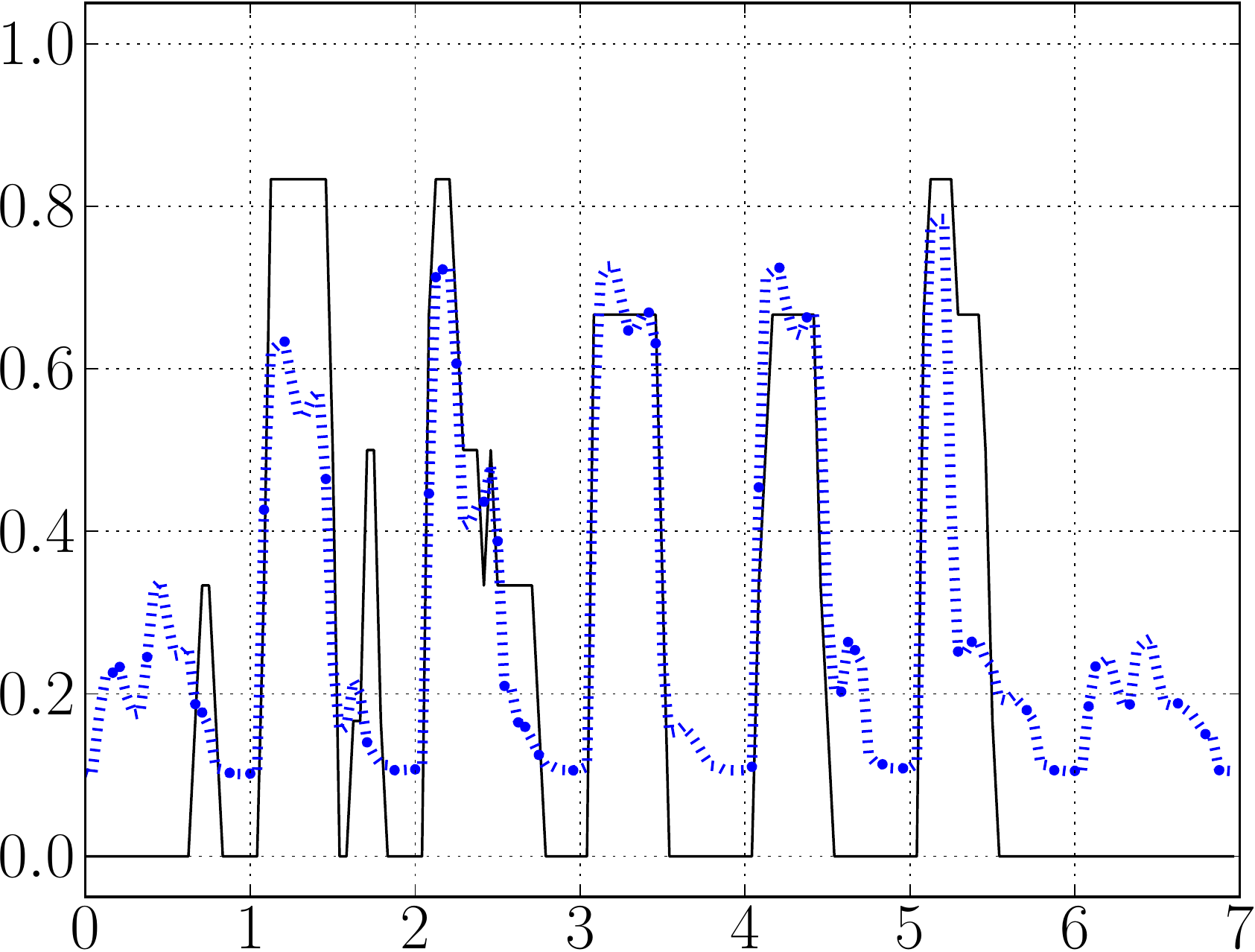} \\
\includegraphics[width=0.46\columnwidth]{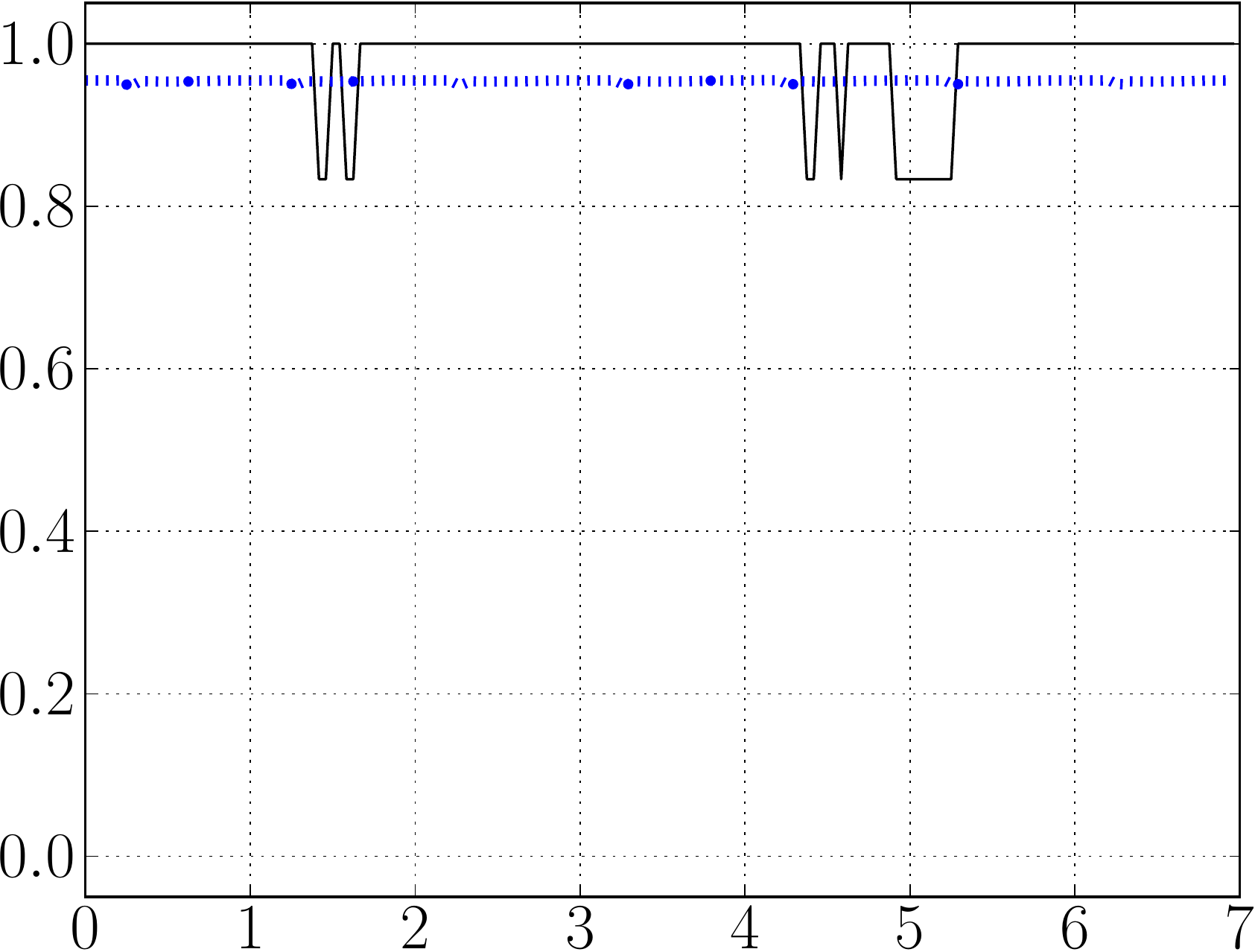} &
\includegraphics[width=0.46\columnwidth]{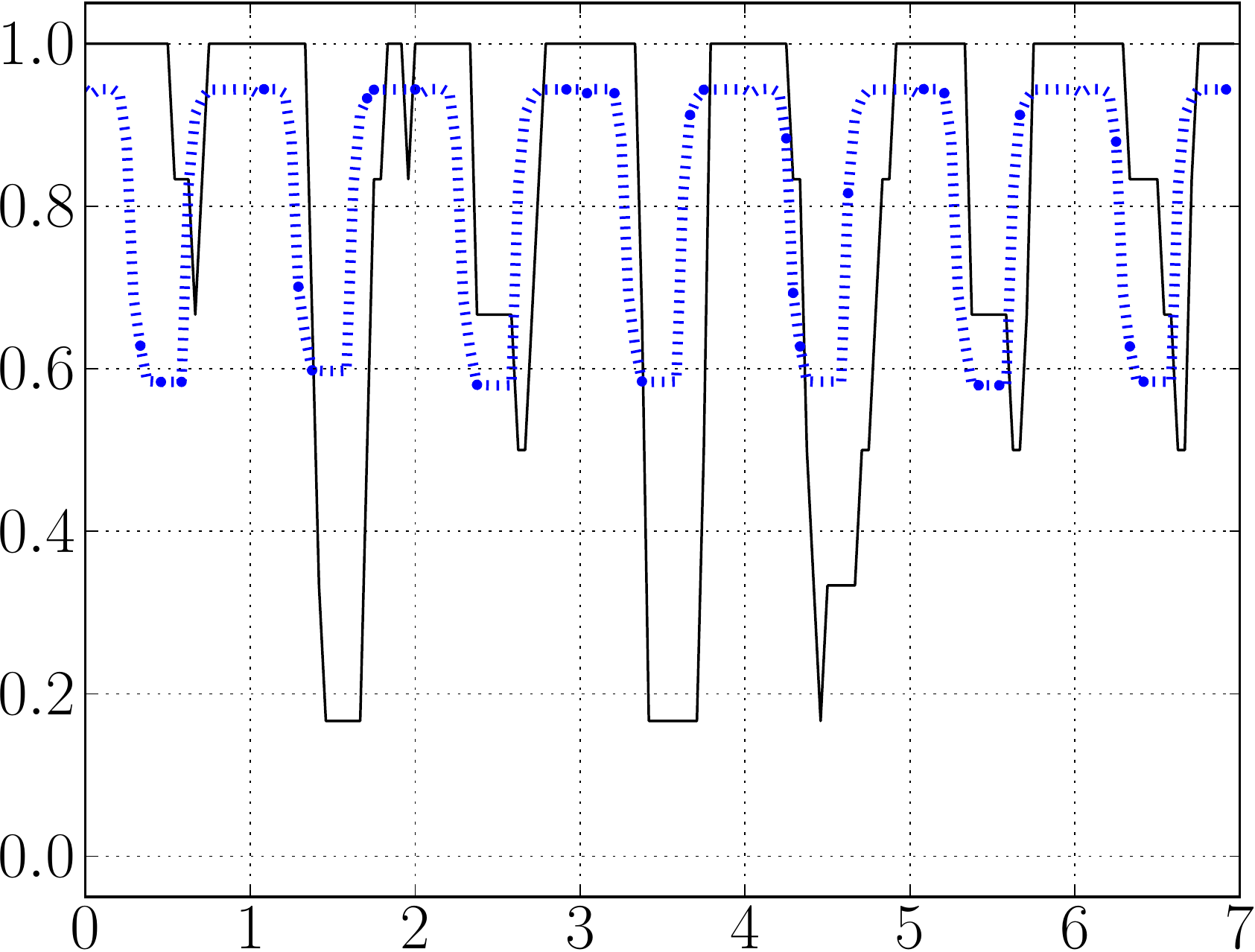} \\
\includegraphics[width=0.46\columnwidth]{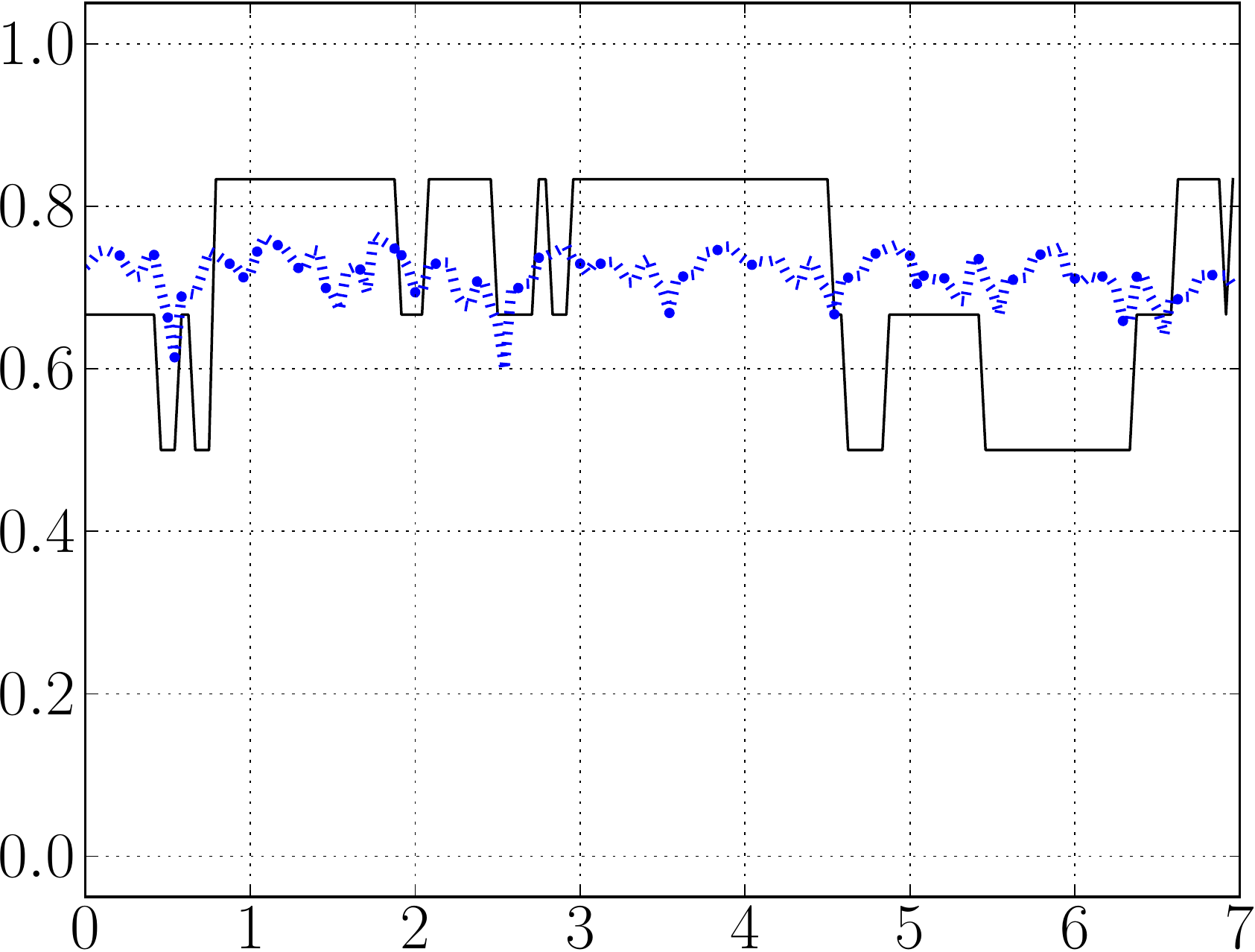} &
\includegraphics[width=0.46\columnwidth]{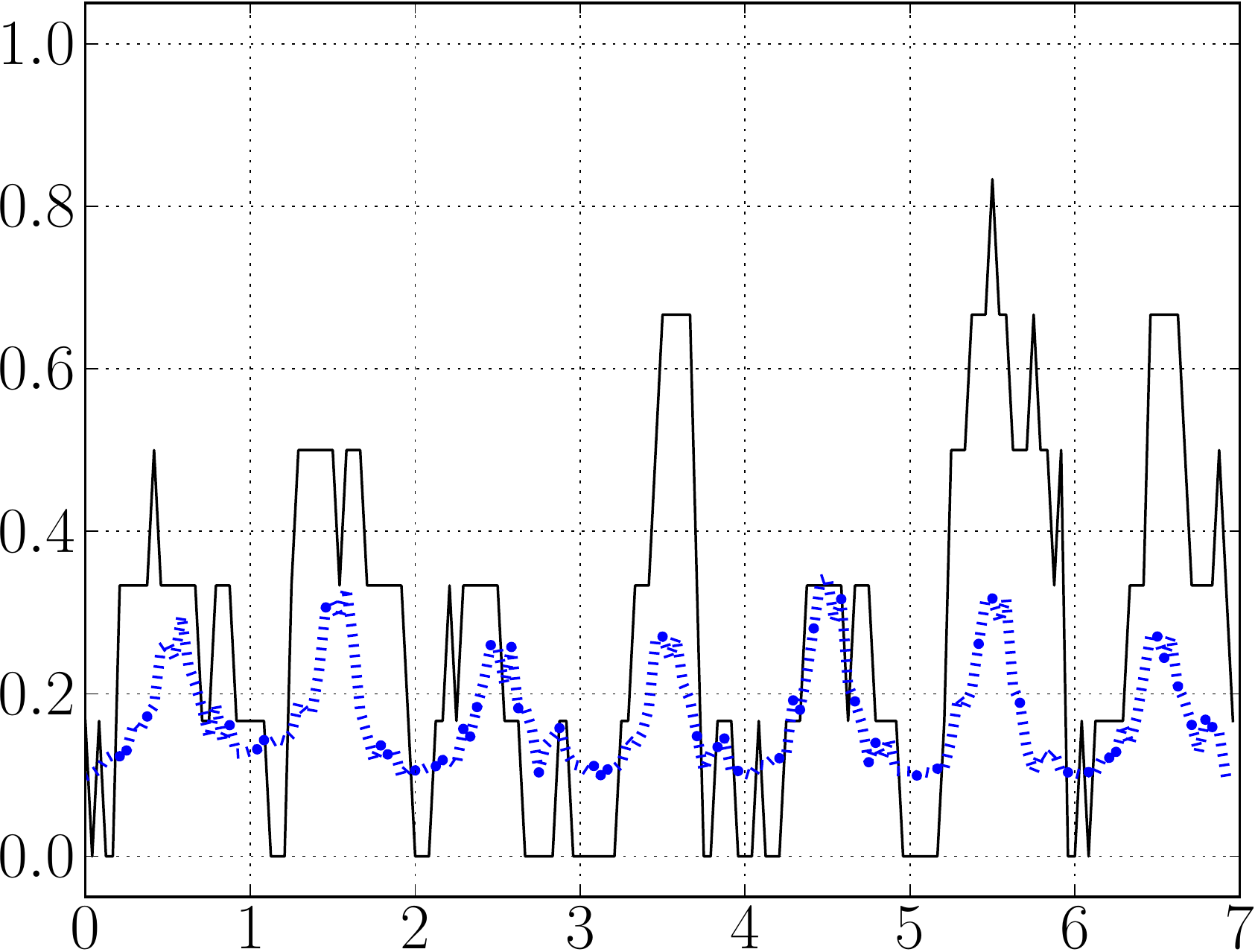} 
\end{tabular}
\end{center}
\caption{Examples of observed availability (solid lines) and predictions (dotted lines) for six representative users (top to bottom: IM, GW, Kad filtered).}
\label{fig:nodes}
\end{figure}

As examples of prediction performance, in Figure~\vref{fig:nodes} we
show predictions (dotted lines) and availability traces (solid lines)
on the six weeks of test data for six typical users, two from each
dataset. \revised{We selected these users to provide concrete examples
  of typical per-user availability patterns in the different datasets
  and of the way our model captures them.}  Since our predictor has a
weekly period, we show a week of predictions and the corresponding
average availability (same time of day and day of week) over the six
weeks of the test set. In IM filtered, nodes frequently have strong
daily periodicity (top left), often accompanied by a different
behavior in the weekend (top right). These features can be easily
recognized by our classifier, and this explains the particularly good
performance of \revised{the predictors using only} the individual
daily and weekly \revised{features}. In GW, nodes are frequently
mostly on, with sporadic disconnections due either to failures or
\rerevised{the} measurement artifacts \rerevised{mentioned in
  Section~\ref{sec:gw}} (center left); however, some users
turn on their gateways only occasionally and mostly during the
day (center right). This explains the good performance of
\revised{predictors using} individual daily and flat
\revised{features}. In Kad filtered, many nodes have availability
patterns with varying daily periodicity (bottom): even though
\revised{predictions using the} ``individual daily'' \revised{feature}
perform best, the confidence level -- as expressed by the GM metric --
is low.

The values of $\betavect$ in Table \ref{tab:accuracy} are expected to
be large when associated to \revised{features} with high predictive
quality, unless some of them convey redundant information. \revised{We
  only observe positive values} for individual \revised{features};
some negative values are instead associated to global
\revised{features}. The net result of this, when associated with
higher positive values on the corresponding individual
\revised{feature}, is a case where the deviations of a node from the
global trend are essentially given more importance and therefore
amplified. In all cases, the standard deviation expressing the level
of uncertainty associated to $\betavect$ is low due to the amount of
data processed, and guarantees that the above considerations on the
relative importance of individual \revised{features} are robust.

\begin{figure}
\begin{centering}
\includegraphics[width=\columnwidth]{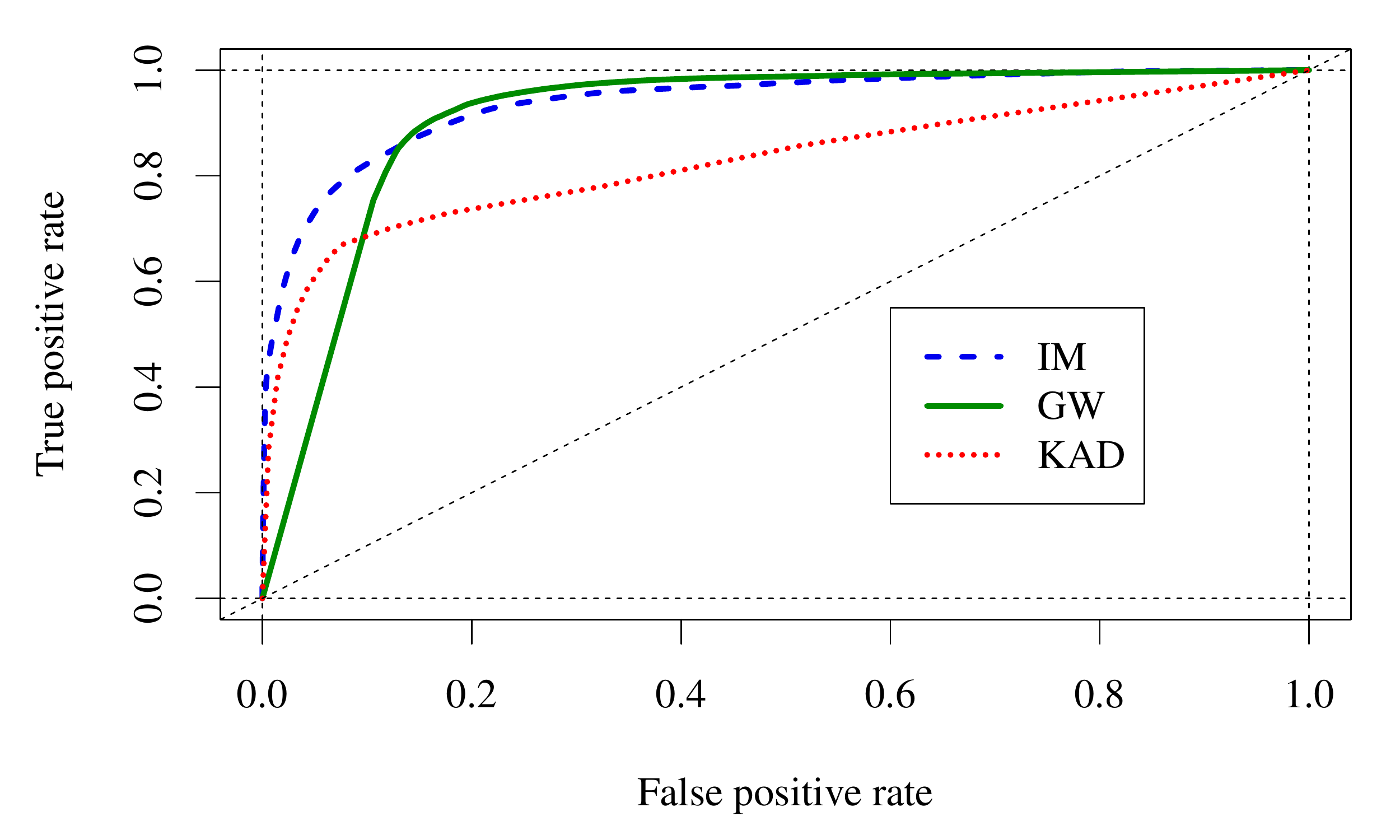}
\end{centering}
\caption{ROC curves corresponding to the logistic regression classifier using a combination of all \revised{features} for \rerevised{our} data sets.}
\label{fig:ROC:curves}
\end{figure}

\rerevised{To understand the achievable false positive/false negative
  rates, we plot in Figure \vref{fig:ROC:curves}} ROC curves for our
non-filtered datasets.  \rerevised{The classifier has a good overall
performance for IM; i}n contrast, it is difficult to
recognize online users in Kad: this is because several users go online
only sporadically and rather unpredictably.  Conversely, in the GW
dataset, for many users it is the downtime which is sporadic and
unpredictable.


\begin{figure}[t]
\centering
\includegraphics[width=\columnwidth]{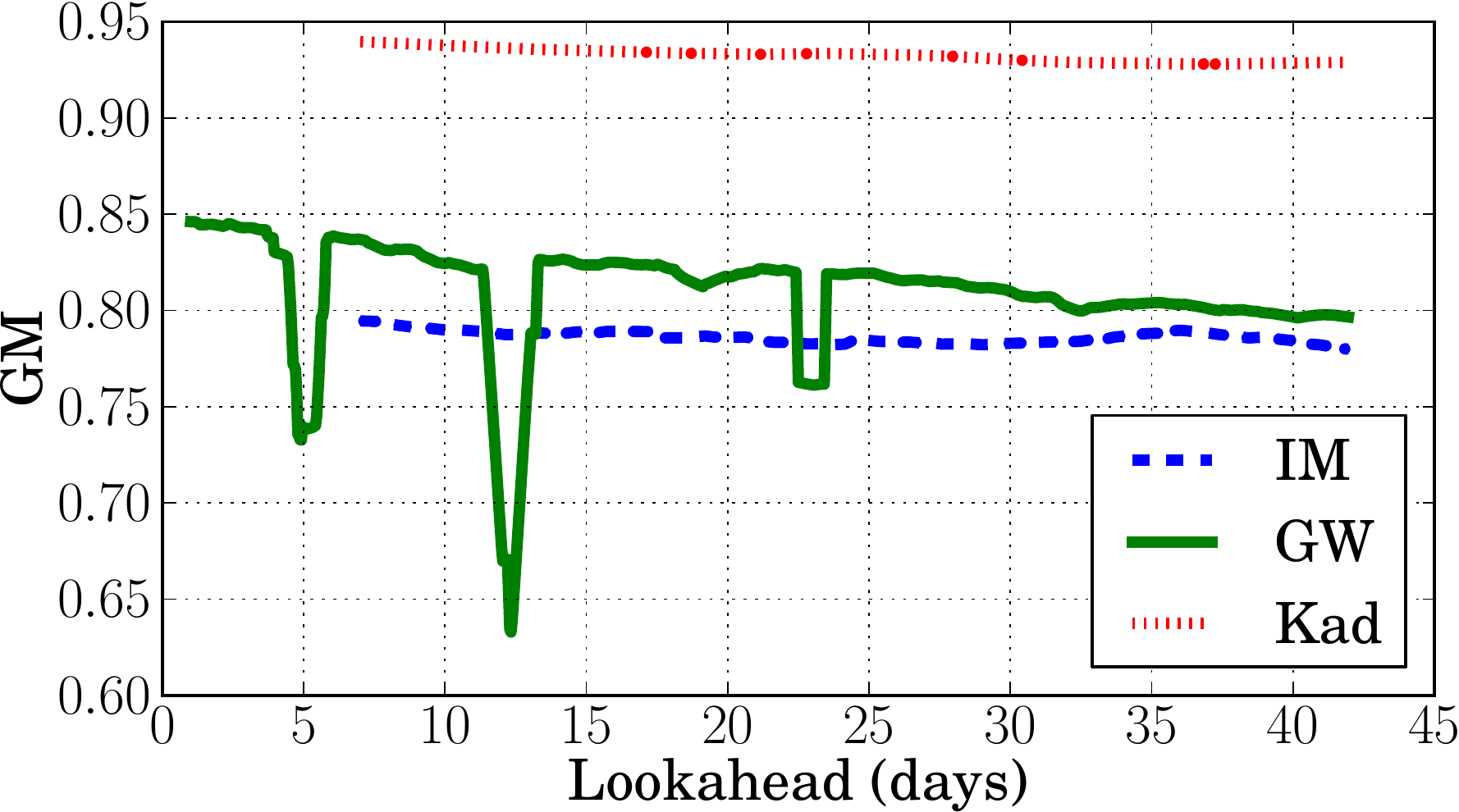}
\caption{GM as a function of time.}
\label{fig:GMvsT}
\end{figure}

In Figure \vref{fig:GMvsT}, we show the evolution of our error metrics
over the test period, averaged on a running window of one week in
order to discount daily and weekly periodic behavior -- except for GW,
where we use a window of one day since the clearly visible measurement
artifacts \rerevised{reported in Section~\ref{sec:gw}} would otherwise
make the plot unreadable. While there is a decreasing trend -- \ie,
prediction quality becomes worse as predictions are made farther into
the future -- this decrease is only marginal: it is, indeed, possible
to predict node availability even on the long run.

To summarize, the main results of this evaluation are two: first,
individual periodic availability patterns enable an effective
prediction of user availability even in the long term; second, the
extent to which user availability is predictable is strongly dependent
on the nature of the application, and in particular to the implicit
and explicit incentives for availability offered by each application.

\subsection{Comparison with Mickens and Noble}
\label{sec:mn}

As discussed in Section~\ref{sec:Related-Work}, the method proposed by
Mickens and Noble (MN hereafter)~\cite{mickens2006exploiting} is the
most closely related piece of work, being the only one that
explicitely tries to predict single nodes' availability.

Before even discussing accuracy, we remark that our method has two
advantages:
\begin{itemize}
\item \emph{Scalability.} Our approach is suitable for large amounts
  of data (\emph{e.g.}, the 1.6 billion data points of the Kad
  dataset). Conversely, MN requires training and generating
  predictions for a set of several predictors for each user and at
  each timeslot; this requires rather expensive operations
  (\emph{e.g.}, matrix inversion) for \emph{each} data
  point. \rerevised{Due to these factors, we found our method to be
    able to process data at a speed which is faster by 4-5 orders of
    magnitude.}
\item \emph{Expressivity.} The output of our method is a
  \emph{probability} of encountering a user online, while MN just
  outputs predictions without a probability. MN's prediction of
  ``online'' (or ``offline'') does not give any estimation over the
  degree of certainty of that prediction. As such, it would be
  impossible to use MN \revised{in the place of our approach in
    \emph{any} of the} applications we consider in
  Section~\ref{sec:Applications}.
\end{itemize}

With these considerations in mind, we proceed describing our
experiments and comparing the prediction accuracy of each approach. We
implemented the MN approach and used it with the parameters
of~\cite[Section~4.1]{mickens2006exploiting}. Due to the limited scalability of the MN method, we create a training set for
both our proposal and MN where user availability is sampled every 6
hours; we train our model as before, while the training period for MN
is the first 18 weeks (corresponding to periods A, B, and C of Section~\ref{sec:Our-Model}).  For the
smaller IM dataset, we evaluate both approaches on all users and bin
them in five availability classes according to the average ratio of
time spent online in the test period:
$[0,0.2)\ldots[0.8,1)$. Similarly, for the larger GW and Kad datasets,
we create a sample of 100 users for each one of the ten availability
class $[0,0.1)\ldots[0.9,1)$. For our approach, we predict that the
user will be online when the predicted probability is greater than
0.5, and offline otherwise.

\begin{figure*}

\centering

\subfloat[IM.]{
        \includegraphics[width=.32\textwidth]{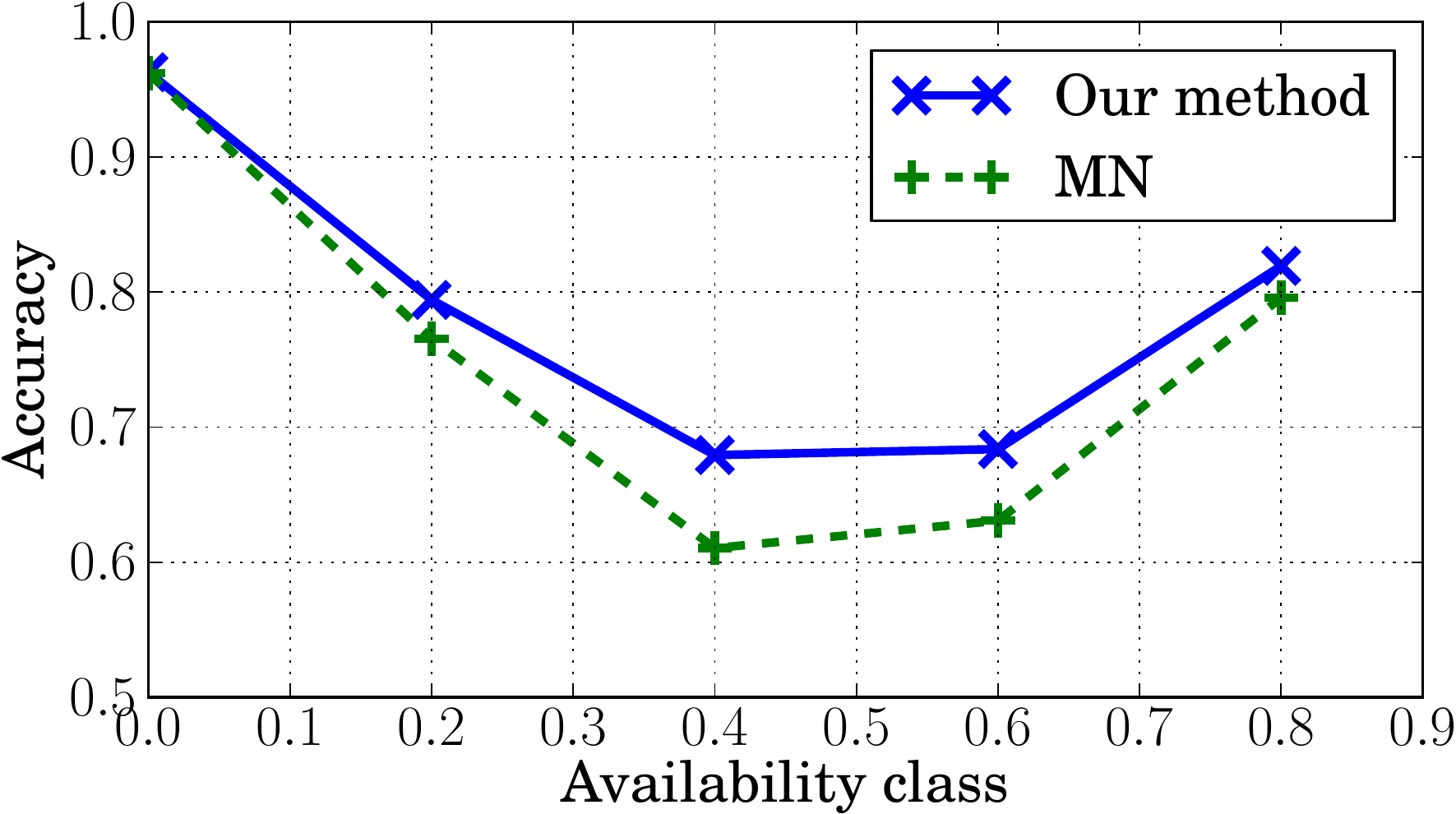} \label{fig:mn-im}
        }
\subfloat[GW.]{
        \includegraphics[width=.32\textwidth]{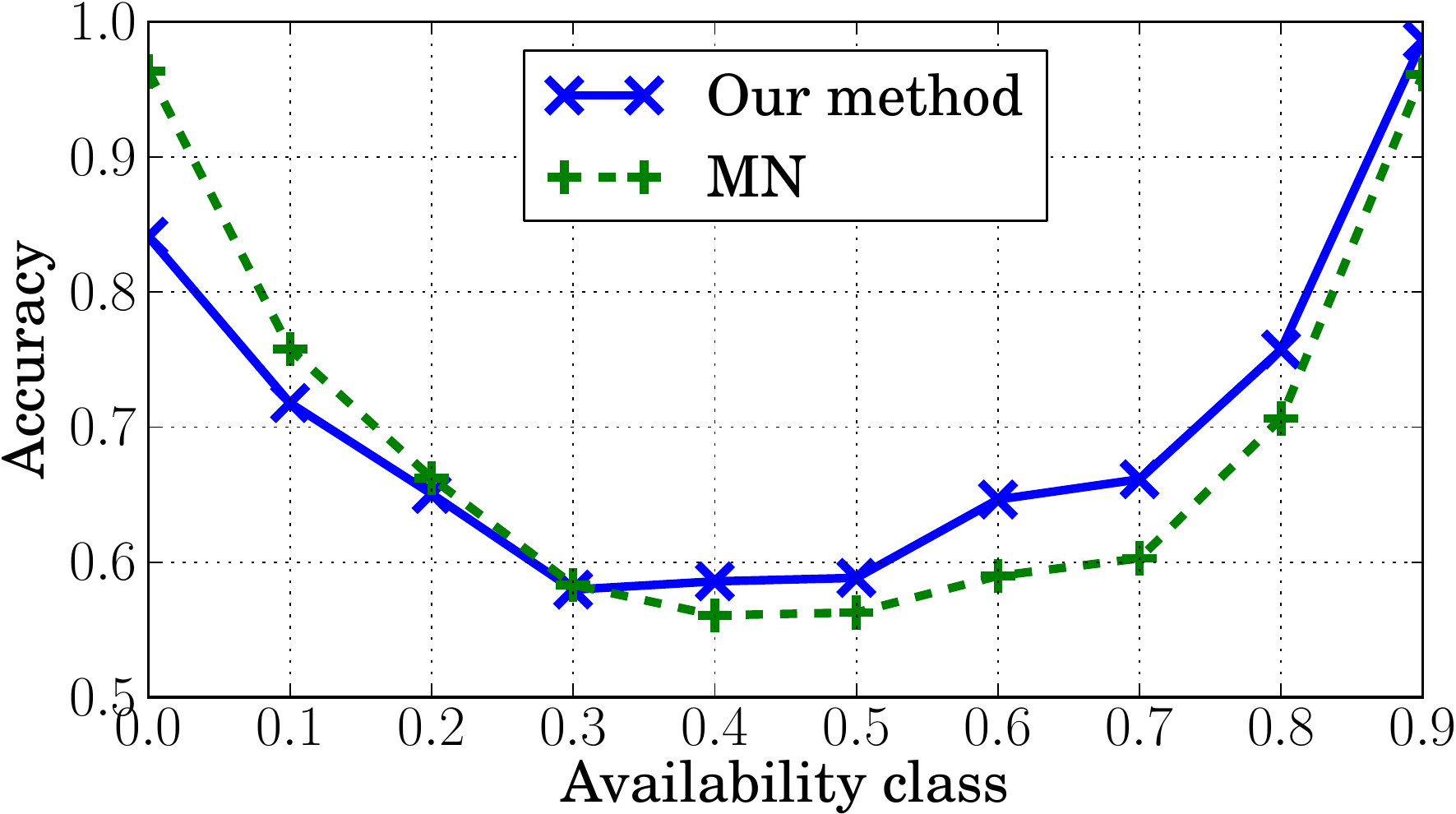}
        \label{fig:mn-gw}
}
\subfloat[Kad.]{
        \includegraphics[width=.32\textwidth]{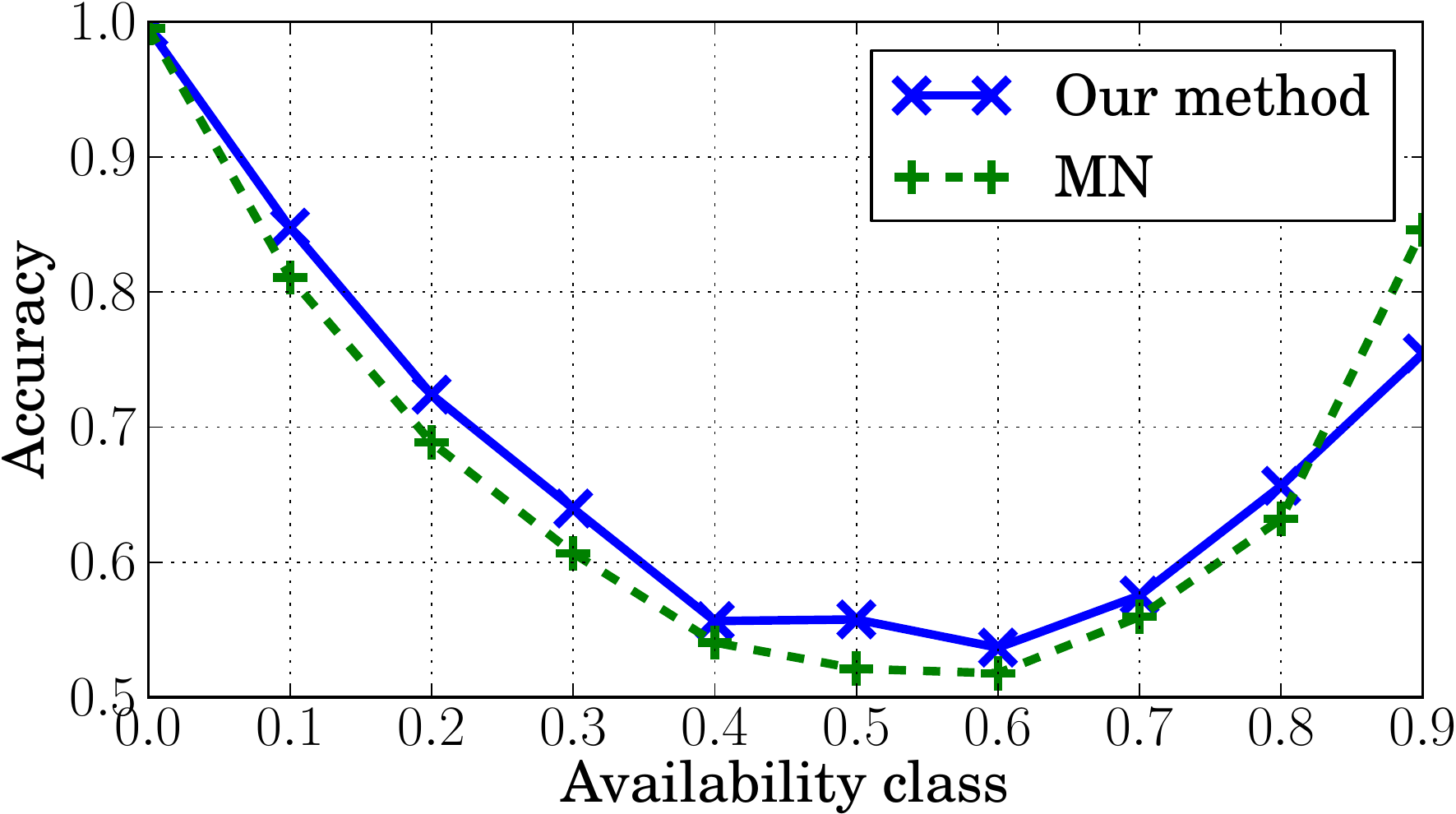}
        \label{fig:mn-kad}
}

\caption{Accuracy comparison with Mickens and Noble's algorithm.}
\label{fig:mn}

\end{figure*}

In Figure~\vref{fig:mn}, we show the per-class accuracy for the
predictions in each each of the availability classes. Besides noting
that extreme (\emph{e.g.}, almost always on or off) cases are easier
to predict in both cases, we note that our method obtains
\rerevised{in most cases} better performance than MN, in particular
for the most frequent instances (\ie, mostly-on nodes in GW and
mostly-off nodes in Kad). It can be noted that both approaches involve
the combination of various \revised{features}; an analysis of the
possible benefits obtained by using \revised{the predictors from MN as
  additional features} is in our plans for future work.

\section{Applications}

\label{sec:Applications}

In this section, we report three examples of application use cases;
they should be regarded as proofs of concept, with the goal of showing
how probabilistic availability prediction can easily and effectively
be integrated in a large class of realistic systems.

\subsection{Node Placement on DHTs}

\label{sub:app-dht}

Distributed Hash Tables (DHTs) are data structures for decentralized
systems that store key/value pairs and provide an efficient lookup
function. We consider a DHT holding a large amount of data (large
values and/or very large quantities of key/value pairs), where
information is stored on a long-term basis. In this case, data is not
erased from nodes between online sessions; hence, data maintenance is
required only when peers abandon the system for good. \revised{In such
  a situation, reaching high availability for data is very expensive
  in terms of resources: it is generally obtained by increasing data
  redundancy, which is an expensive strategy because it entails
  increased usage of both bandwidth and storage space on peers. Here,
  we show how a simple data placement policy informed by our
  predictions can result in higher data availability without requiring
  increased resource usage.}

Mickens and Noble also proposed using availability prediction to
optimize DHTs \cite{mickens2006exploiting}. They considered a case
where each (transient) disconnection triggers data maintenance, and
proposed to alleviate the maintenance load by storing data only on the
most available peers. This approach has the obvious weakness of
overloading the most stable nodes, providing them an incentive to
lower their availability. Differently from such approach, we focus on
improving data availability without imposing any additional burden on
any peer.

As in Chord \cite{stoica2003chord}, persistent node identifiers and
hash values for keys are placed on a logical ring topology, and each
key/value pair is replicated on a \emph{neighbor set} of $n$ nodes,
whose identifiers are the closest successors to the hash value of
the key in the ring.

In general, node identifiers in DHTs are chosen randomly.  We propose
instead a smart policy that maximizes data availability,
\ie the fraction of time slots in which at least one peer in a
neighbor set is available. For example, it would be wise to distribute
two replicas of the same piece of data on a node which is often
available during the day and another which is available at night.

For a given neighbor set $N$ and a set of time slots $T$, we compute
the predicted data availability (\emph{i.e.}, the probability that
at least one node holding a replica is online) as a function of the
prediction $P$:
\begin{equation}
\frac{\sum_{t\in T}1-\prod_{n\in N}\left(1-P_{n,t}\right)}{\left|T\right|}.\label{eq:dht-data-avail}
\end{equation}
 It is important to note that, by using the above formula, we are
assuming that the probabilities of being available for two nodes in
the same timeslot are independent. This assumption is not true when,
for example, two nodes are disconnected at the same time for the same
reason, such as a network outage or an external event, so this might
introduce discrepancies between the predicted and observed availability
distributions.







\begin{algorithm}
\textbf{Input}: prediction matrix $P$

\textbf{Output}: mapping $M$ from nodes to identifiers
\begin{lyxcode}
\textbf{function~$PA(n,M)$:}

\textbf{~~return~}\textrm{average~availability~on~neighbor}

~~~~\textrm{sets~of~$n$~using~Equation~\ref{eq:dht-data-avail}}

~

$N\leftarrow1,000$~\textrm{\#~number~of~iterations}

$M\leftarrow$\textrm{mapping~with~a~random~id~per~node}

\textbf{for}~$i\in1,\ldots,N$\textbf{:}

\textbf{~~for~each~}\textrm{node}~$n$:

~~~~\textrm{choose~a~random~node}~$n'$

~~~~$a_{0}\leftarrow PA(n,M)+PA(n',M)$

~~~~$M'\leftarrow M$\textrm{~switching~the~ids~of~$n$~and~$n'$}

~~~~$a_{1}\leftarrow PA(n,M')+PA(n',M')$

~~~~\textbf{if~}$a_{0}<a_{1}$:~$M\leftarrow M'$

return~$M$
\end{lyxcode}
\caption{\label{alg:Node-identifier-assignment}Node identifier assignment
in a DHT.}
\end{algorithm}

Our procedure for assigning node identifiers
(Algorithm~\vref{alg:Node-identifier-assignment}) works by starting
with a randomly assigned identifier per node, iteratively considering
for each node another random one, and verifying (based on Equation
\ref{eq:dht-data-avail}) whether exchanging their identifiers would
enhance, on average, the \revised{average} predicted data availability
for the involved neighbor sets. If so, their identifiers (and hence
positions in the ring) are exchanged.

This policy can be easily implemented in a decentralized network, if
each node starts with an availability prediction for itself. Any node
can easily find candidates for exchanging positions by performing a
lookup of a random key in the DHT; the availability prediction
according to Equation \ref{eq:dht-data-avail} can be computed if each
node shares its availability predictions with neighbors. To avoid
performing too many expensive \rerevised{data exchanges}, nodes can
start accepting data once their identifier is sufficiently stable. In
addition, this method can \rerevised{adapt} over time as new nodes
join and predictions get updated by continuing to look for candidates
for switching positions. We assume that nodes correctly follow the
protocol specified here: the design of detection and incentive
mechanisms is outside the scope of this work.

\paragraph*{Experiments}

In this use case, it is recommendable to only store data on nodes with
reasonably high availability\rerevised{: b}ecause of this, we
\rerevised{performed} these experiments on the three filtered
datasets. We trained our model on the A, B and C periods, and we
executed Algorithm~\ref{alg:Node-identifier-assignment} to assign node
identifiers based on the \rerevised{D period predictions}. We
\rerevised{then} measured predicted and simulated data availability
according to the data placement outcome, repeating each instance of
the experiment 100 times. The smallest filtered dataset is IM, with
408 users in the C and D periods. \rerevised{To} have experiments of
the same \rerevised{size}, we chose a random sample of 408 nodes
\rerevised{for} GW and Kad in each instance of the experiment. We set
\textit{a priori} the \rerevised{neighbor set size} $n$ based on the
average availability $\overline{a}$ of nodes in period C (0.488 for
IM, 0.377 for Kad and 0.939 for GW), choosing the smallest $n$
\rerevised{satisfying} $\left(1-\overline{a}\right)^{n}<0.01$.  This
corresponds to a simple uniform availability model \rerevised{to set}
redundancy aiming for similar starting conditions in the three
datasets, and it results in $n=2$ for GW, $n=7$ for IM, and $n=10$ for
Kad.

We note that all three datasets can be representative of systems where
a DHT is run: an instant messaging application to look up data about
users, a file-sharing application, or a P2P application running on
high-availability set-top-boxes~\cite{nada}.

\begin{table}
\begin{centering}
\begin{tabular}{lrrr}
\hline 
\multirow{1}{*}{Dataset} & \multicolumn{1}{c}{Random} & \multicolumn{1}{c}{Prediction-based} & \multicolumn{1}{c}{$\rho$} \tabularnewline
\hline 
IM & $97.86\%\pm0.51$ {[}0.77{]} & $\mathbf{99.95}\pmb\%\pm0.06$ {[}0.11{]} & 392\%\tabularnewline
GW & $99.18\%\pm0.26$ {[}0.20{]} & $\mathbf{99.54}\pmb\%\pm0.13$ {[}0.12{]} & 12\%\tabularnewline
Kad & $95.17\%\pm0.86$ {[}0.77{]} & $\mathbf{96.30}\pmb\%\pm0.86$ {[}0.76{]} & 37\%\tabularnewline
\hline 
\end{tabular}
\par\end{centering}

\caption{Data availability for the DHT simulation.}
\label{table:dht}
\end{table}

In Table~\vref{table:dht} we show predicted and real data availability
(\emph{i.e.}, the measured \revised{percentage} of time slots in which
at least a replica is online, averaged over each neighbor set) on the
test period D (see Section \ref{sec:Our-Model}). \rerevised{We report
  the average availability and standard deviation between instances of
  the experiment. In square brackets, we show the average difference
  between predicted and real data availability for the whole system.}

\rerevised{From Table~\ref{table:dht}}, we note that
our optimized data placement strategy consistently performs better
than the standard random placement of nodes in a DHT; however, the
increase in availability varies between the different datasets: it is
most significant in IM, where the average time that a piece of data is
unavailable drops from around 3.5 hours to five minutes per week; it
is instead less impressive in Kad, where unavailability decreases from
around eight to six hours per week.  We attribute this to the
difference between the datasets: when behavior is strongly periodic
and easy to predict, optimized data placement results in much better
availability. However, a decrease in unavailability by around a
quarter is still desirable, in particular because it is obtained
without any extra requirement on any node.

Reaching high availability in online storage is very expensive. Using
the common availability model that assumes homogeneous storage nodes
and independent failures (\emph{e.g.},~\cite{bhagwan-et-al-04}) the
required redundancy for replicated storage is directly proportional to
the ``number of nines'' for required data availability. More formally,
in that model data availability is computed as
$a = 1 - (1-a_n)^r$, where $a_n$ is the availability of any storage
node and $r$ is the replication factor. It is easy to
derive~\cite{castro05} that in order to increase data availability
from $a_0$ to $a_1$, the replication factor has to increase by a ratio
of
$$
\rho = \frac{r_1 - r_0}{r_0} = \frac{\log (1-a_1)}{\log (1-a_0)} - 1.
$$ In order to help interpreting the results of our evaluation,
\rerevised{in Table~\ref{table:dht}} we provide the values of $\rho$,
which is determined solely by the availability values observed in our
trace-driven simulations, and can be interpreted as an ``equivalent
redundancy increase'': \emph{i.e.}, the ratio of additional redundancy
\revised{(and corresponding overhead in terms of bandwidth and storage
  space)} that should be injected in the system in order to obtain an
equivalent increase in availability. The improvement in IM can be
considered as equivalent to an almost 400\% increase in redundancy;
even in the less impressive GW and Kad cases, the value of $\rho$ is
far from negligible in terms of resource economy.

Based on the numbers in square brackets in Table~\ref{table:dht}, we
also notice that there is a reasonable match between predicted and
real availability values. This would not happen if there were strong
dependencies between availabilities of different nodes (\emph{e.g.},
two users disconnecting at the same time because of the same external
event); as such, we believe the hypothesis of independence we used in
Equation \ref{eq:dht-data-avail} to be sensible.

\subsection{Data Placement for F2F Storage}\label{sub:app-storage}

\rerevised{ We now show that our probabilistic classifier can be used
  to drive data placement in other declinations of P2P storage. We now
  consider a use case where data is still stored on users' machines,
  and we still want to optimize data availability by ensuring that at
  least one replica of the data is online; however, we have more
  constraints on data placement.}

Friend-to-friend (F2F) storage \rerevised{is an interesting instance
  of P2P storage}. In these systems, nodes are constrained to store
data only on machines owned by {}``friends'', with reciprocal
real-world trust bonds. These trust relationships can be leveraged to
obtain guarantees of dependability, privacy and security, since
trusted users are unlikely to behave maliciously, selfishly or anyway
deviating from the expected behavior
\cite{cutillo2009safebook,tran2008friendstore}.

\begin{algorithm}[t]
\textbf{Input}: prediction matrix $P$, per-node capacity $k$

\textbf{Output}: mapping $M$ such that $p_{1}\in M(p_{2})$ if $p_{1}$
holds data for $p_{2}$
\begin{lyxcode}
\textbf{function~$PA(F)$:}

\textbf{~~return~}\textrm{predicted~availability}

~~~~\textrm{for~data~stored~in~friend~set~$F$}

~

\textbf{function~$\Delta\left(F,n\right)$:}

~~\#\textrm{~increase~in~availability~due~to~$n$}

~~\textbf{return~}$PA\left(F\cup\left\{ n\right\} \right)-PA\left(F\setminus\left\{ n\right\} \right)$

~

\#\textrm{~random~initialization~of~}$M$

\textbf{for~each}\textrm{~node~}$n$:

~~$F\leftarrow k$\textrm{~random~friends~of~}$n$

~~\textbf{for~each}~$f\in F$:

~~~~$M(f)\leftarrow M(f)\cap\{n\}$

\#\textrm{~optimization}

\textbf{while~true:}

~~\textbf{$c\leftarrow$false~}\#\textrm{~have~changes~been~made?}

\textbf{~~for~each~}\textrm{node}~$n$:

~~~\textbf{~}$F_{0}\leftarrow$\textrm{~friends~of~$n$~such~that~$n\in M(f)$}

~~~~$f_{0}\leftarrow\arg\min_{f\in F_{0}}\Delta\left(M\left(f\right),n\right)$

~~\textbf{~~}$F_{1}\leftarrow$\textrm{~friends~of~$n$~such~that~$n\notin M(f)$}

~~~~$f_{1}\leftarrow\arg\max_{f\in F_{1}}\Delta\left(M\left(f\right),n\right)$

~~~~\textbf{if~$\Delta\left(M\left(f_{0}\right),n\right)<\Delta\left(M\left(f_{1}\right),n\right)$:}

~~~~~~$c\leftarrow$\textbf{true}

~~~~~~$M\left(f_{0}\right)\leftarrow M\left(f_{0}\right)\setminus\left\{ n\right\} $

~~~~~~$M\left(f_{1}\right)\leftarrow M\left(f_{1}\right)\cup\left\{ n\right\} $~~

~~\textbf{if~not~$c$}:\textbf{~return~$M$}
\end{lyxcode}
\caption{\label{alg:Data-placement-strategy}Data placement for F2F storage.}
\end{algorithm}

In our simplified simulation, each node has to store a data object
of unitary size on friend nodes, and has a storage capacity of $k$
units, which is not sufficient to store the data for all the friends.
Similarly to what we propose for DHTs in Section \ref{sub:app-dht},
we optimize the choice of nodes for which to store data in order to
enhance the overall data availability. The predicted data availability
can be computed again from Equation \vref{eq:dht-data-avail}, using
the set of friends storing replicas of a nodes' data object as the
neighbor set $N$.

Algorithm \vref{alg:Data-placement-strategy} optimizes data
availability as follows: each node starts by storing the pieces of
data of a random subset of $n$ friends; afterwards, iteratively, each
node considers erasing the piece of data that would impact less its
predicted availability, and exchanging it with the one that would
improve most its predicted availability. If the net effect on system
availability is positive, then the switch is performed. The algorithm
continues until it stabilizes, when no pieces of data are switched
anymore.

Assuming once again that nodes have access to their own availability
predictions and share them with friends, this algorithm is simple
to implement in a decentralized system. In this case, some level of
security is guaranteed by the fact that all communication happens
among trusted peers. Exchanges of data can remain virtual
until the algorithm has reached convergence, sparing the cost of needless
uploads of data that will be subsequently erased.

Kermarrec \etal have proposed an availability-aware data
placement policy for peer-to-peer storage systems called {}``R\&A''
(Random \& Anti-correlated) \cite{KERMARREC:2010:HAL-00521034:2},
which aims to store data replicas on sets of nodes whose availability
is pairwise anti-correlated. We implemented that policy in our system
with constraints on resources and on eligible data holders (Algorithm
\vref{alg:R+A-algorithm-for}).

\begin{algorithm}
\textbf{Input}: availability matrix $A$, per-node capacity $k$

\textbf{Output}: mapping $M$ such that $p_{1}\in M(p_{2})$ if $p_{1}$
holds data for $p_{2}$
\begin{lyxcode}
\textbf{function~$C\left(n_{1},n_{2}\right)$:}\textrm{~\#~correlation}

~~\textbf{return~$\sum_{t\in T}(1\,\mathrm{if}\, A_{n_{1},t}=A_{n_{2},t},\,\mathrm{else}\,0)$}

~~

\textrm{\#~free~storage~per~node}

$S\leftarrow$\textrm{mapping~s.t.~$S(n)=k$~for~each~node~$n$}

\textbf{while}~$\sum_{n}S(n)>0$:

~~\textbf{for~each~}\textrm{node}~$n$:

~~~~$F\leftarrow$\textrm{friends~of~$n$~s.t.~$S(r)>0$}

~~~~\textbf{if~}$\left|F\right|=0$:~\textrm{skip~to~next~$n$}

~~~~$r\leftarrow$\textrm{random~element~of~$F$}

~~~~$M(n)\leftarrow M\left(n\right)\cup\left\{ r\right\} $;$S(r)\leftarrow S\left(r\right)-1$

~~~~\textbf{if~}$\left|F\right|=1$:~\textrm{skip~to~next~$n$}

~~~~$a\leftarrow\arg\min_{f\in F}\sum_{t\in T}C(r,f)$

~~~~$M(n)\leftarrow M\left(n\right)\cup\left\{ a\right\} $;~$S(a)\leftarrow S\left(a\right)-1$

\textbf{return~$M$}
\end{lyxcode}
\caption{\label{alg:R+A-algorithm-for}R\&A algorithm for F2F storage.}
\end{algorithm}

\paragraph*{Experiments}

\begin{table}
\begin{centering}
\begin{tabular}{lrrr}
\hline 
\multirow{1}{*}{Dataset} & \multicolumn{1}{c}{R\&A} & \multicolumn{1}{c}{Prediction-based} & $\rho$\tabularnewline
\hline 
IM & $99.64\%\pm0.09$ & $\mathbf{99.90}\pmb\%\pm0.04$ {[}0.04{]} & 23\%\tabularnewline
GW & $97.09\%\pm0.52$ & $\mathbf{99.45}\pmb\%\pm0.17$ {[}0.13{]} & 47\%\tabularnewline
Kad & $95.89\%\pm0.74$ & $\mathbf{96.34}\pmb\%\pm0.91$ {[}0.67{]} & 4\%\tabularnewline
\hline 
\end{tabular}
\par\end{centering}

\caption{Data availability for the F2F simulation.}
\label{tab:storage}
\end{table}

In order to have simulation results that are comparable with the DHT
experiments, we have defined parameters that are analogous
to the ones described in the previous section. We adopted
the three filtered datasets, and samples of 408 nodes for each
run of the experiment; the $k$ value describing the storage space on
each node is set analogously to the redundancy value for the DHT
simulation ($k=7$ for IM, $k=2$ for GW, and $k=10$ for Kad). To create
a synthetic model of a social network, we generated a network
according to the Watts-Strogatz small-world model
\cite{watts1998collective}, with average degree (\emph{i.e.}, number
of friends per user) 20 and rewiring probability 0.5.

The results on data availability for this setting are reported in
Table \vref{tab:storage}. By comparing it with Table \vref{table:dht},
we note that without the proposed approach we obtain a slightly lower
availability with respect to the DHT after optimization; we attribute
this to the lower flexibility in this scenario: while a node can take
any position in the DHT, it can only store data for friends in the F2F
case.

The R\&A placement strategy suffers from the fact that this approach
does not allow re-allocating data when poor choices have already been
made, in particular for the GW case where a few pieces of data are
replicated 0 or 1 times. By adopting a more accurate model of data
availability and by reclaiming already allocated space in order to use
it better, our proposal obtains noticeably better performance.

\paragraph*{Discussion}

The two scenarios of DHT node placement and F2F data placement bear
strong similarities with each other. In both cases data availability
can be modeled as a function of node availability and a design choice
(respectively, choice of node identifiers and data placement). Probabilistic
availability predictions make it easy to create cost functions such
as Equation \ref{eq:dht-data-avail}, which is fed to optimization
algorithms. We point out that such an approach can be extended in
several ways, including for example erasure coding, plus different
application settings and storage constraint variants.

We evaluated \emph{local }optimization algorithms, which optimize
for predicted availability by changing only part of the solution at
a time; they can get stuck in local optima. The study and evaluation
of more elaborate techniques to find better solutions is outside the
scope of this work, and a direction for further research.

\subsection{Newsfeed Pre-Loading}

\label{sub:sn-cache-prefetching}

\rerevised{We now move away from the peer-to-peer use cases, and
  consider a case in which we predict the availability of users
  behaving as clients; in this case, we use
  availability prediction to drive pre-loading of data for users that
  are most likely to connect shortly.}

On social networking websites, users are often presented with some
sort of {}``newsfeed'' personalized page including recent updates by
friends or followed contacts. Generating such pages involves serious
scalability problems, since -- due to the sheer size of databases
involved -- information about contacts is scattered across a large
amount of servers in a data center; creating a newsfeed page requires
obtaining data from a large fraction thereof. Hence, two approaches
have been discussed~\cite{hoff2009pushpull} to address this issue:
\emph{pull-on-demand} and \emph{push-on-change}.

The pull-on-demand approach queries all the relevant
\rerevised{servers} when the user requests a page: this approach
requires pulling information from a large number of \rerevised{servers}
concurrently, which may result in congestion and ultimately high
latency in returning the result. \rerevised{It could be possible to
  mitigate this problem with}
an additional layer of caching, \rerevised{which would} however be an
expensive solution \rerevised{having} cache eviction
problems~\cite{hoff2009pushpull}.

Alternatively, with the push-on-change approach, a set of \rerevised{servers}
responsible for serving newsfeeds are sent user updates when they are
generated. With push-on-change, information is already in a single
place when the newsfeed is requested, making it more efficient to
serve. Clearly, this approach entails data duplication, since each
update generated by a single user will be sent to the \rerevised{servers}
responsible for each of its contacts. With push-on-change, newsfeeds are
essentially pre-loaded for all users, whether or not they access the
system.

Without the pretense of providing a complete system design, we sketch
a hybrid solution combining both behaviors. Pull on demand is the
default, as many users will connect sparingly to the system and
generate few, albeit expensive, requests\rerevised{;} the
push-on-change approach is enabled for the periods of time in which a
user is predicted to be more likely to connect. This approach allows
to increase efficiencies for the most predictable accesses, while
limiting the data duplication that would be generated by enabling
push-on-change for all users.

Since the predicted availability for users varies over time, the set
of users who are enabled for push-on-change uploads also varies; when
a user becomes enabled for profile pre-loading, old updates by contacts
will also need to be sent to them.

\paragraph*{Experiments}

\begin{figure}
\begin{centering}
\includegraphics[width=\columnwidth]{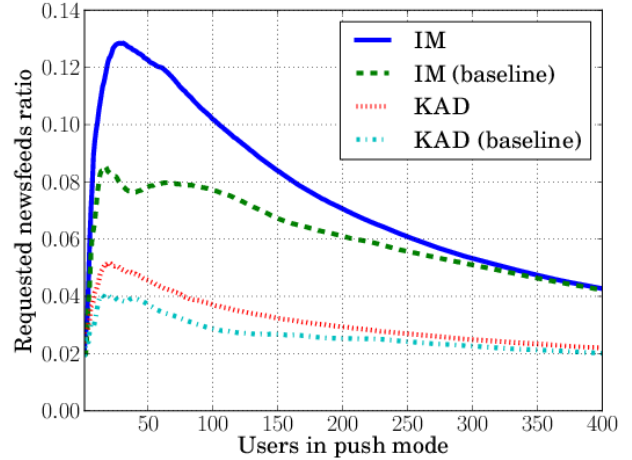}
\par\end{centering}

\caption{Ratio of requested newsfeeds.}
\label{fig:newsfeeds}
\end{figure}

We consider a system that enables push-on-change mode towards $n$
users that are not connected at the moment. For each timeslot in the
IM and Kad traces (where there are frequent connections and
disconnections in user traces), we \rerevised{enable} push-on-change
for the $n$ \rerevised{disconnected} users that are predicted to be
most likely to connect in the following one. A higher availability
ratio for them in the subsequent timeslot corresponds to better
performance for the system. \rerevised{We compare this to a baseline
  approach} that constantly enables push mode for the $n$
\rerevised{disconnected} users that were most available in the
training period C.

The prediction-based approach, by incorporating periodic availability
patterns in the choice, outperforms the baseline; the difference is higher in the IM dataset -- where user behavior
is more regular -- than in the Kad dataset. \rerevised{The}
first few users chosen by both approaches actually have a low probability
of coming online. This is because users with very high predicted availability
that are offline are likely to be always-on nodes that get disconnected
for rather long periods of time, probably due to failures rather than
ordinary user behavior. This would be taken into account by an approach
that differentiates between ordinary and extraordinary downtimes which
is a possible topic for further work.

\section{Conclusion}

\label{sec:Conclusions}

Motivated by the need of a range of Internet applications to address
the problem of resource provisioning, the main focus of our work was
to design an efficient model to produce probabilistic, long-term
forecasts of user uptime. Having realized that user connectivity
patterns are regular and correlated, we built a range
of \revised{features} and combined them with logistic regression, to
estimate the probability for each individual user to be online in a
future time instant. By adopting a probabilistic treatment of logistic
regression, we were also able to assess the importance of individual
\revised{features} in predicting user availability.

We have shown that our method is scalable and we validated it with
trace-driven experiments for three different Internet applications.
In particular, we assessed our predictions on six-month long datasets
comprising a large number of users, in order to validate our methods
on real scenarios.  Our study shows that \rerevised{prediction
quality} is variable across datasets, and that the best results were
obtained for user-traces characterized by strong periodic
behavior. Finally, we demonstrate that the prediction accuracy of our
model fit our goals: \rerevised{prediction quality} decreases only slightly in
the long term.

In Section~\ref{sec:Applications} we also presented three
representative application scenarios and we showed that they all
consistently benefit from the ability to predict user online
availability. Using simple probabilistic models and trace-driven
numerical simulations, we obtained substantial improvements in terms
of the most important performance metrics of each application
domain. Essentially, we showed that current Internet applications can
incorporate and benefit from the predictions of future user
availability with little or no cost.

\revised{
We focused on availability prediction because traces were available to
evaluate it; there are, however, other types of user behavior that
are very interesting to
predict. Scellato \etal~\cite{scellato2011track} and
Traverso \etal~\cite{traverso2012tailgate} show that social and
location-based features can effectively be used to drive caching and
pre-fetching for video serving systems.  We conjecture that a
predictive and probabilistic approach similar to ours could be
applicable to a larger class of use cases where other types of user
behavior in addition to connectivity are taken in
consideration.  }

In conclusion, this work shows that the availability of users is
predictable to a large extent and with good accuracy, and that
applications might benefit from anticipating, rather than reacting to
user demand. Human-generated workloads go beyond simple intermittent
availability patterns: we believe that the ability to predict what a
user will do with an application, in addition to when, is a
challenging and useful topic that can enable better system design.

\bibliographystyle{IEEEtran}
\bibliography{biblio,filippone3}

\vfill

\begin{IEEEbiography}[{\includegraphics[width=1in,height=1.25in,clip,keepaspectratio]{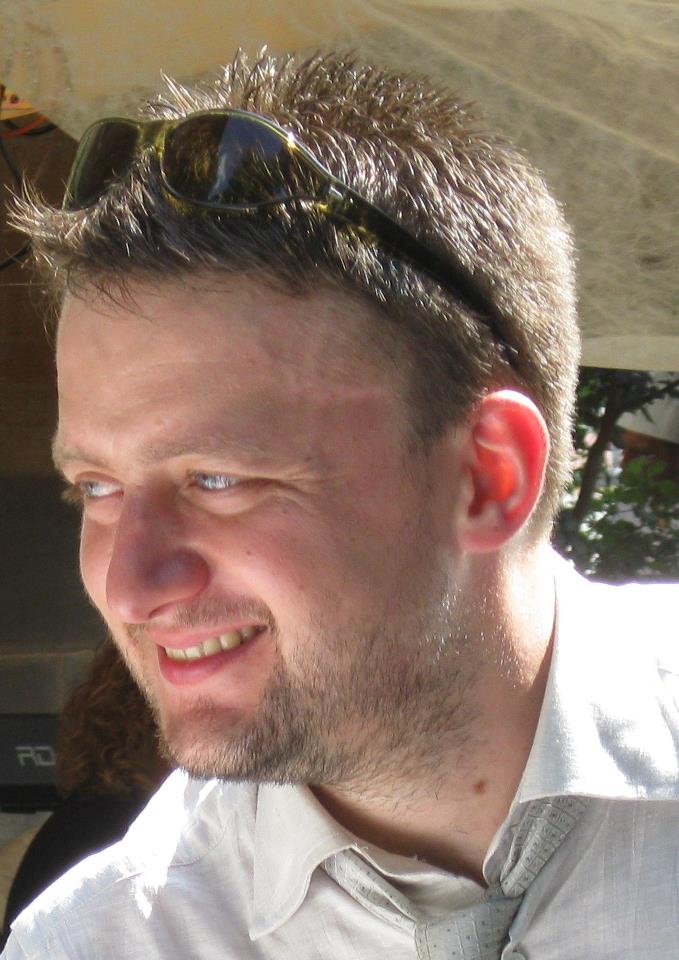}}]{Matteo Dell'Amico}
is a researcher at EURECOM; his research revolves on the topic of
distributed computing.  He received his M.S. (2004) and Ph.D. (2008)
in Computer Science from the University of Genoa (Italy); during his
Ph.D. he also worked at University College London. His research
interests include data-intensive scalable computing, peer-to-peer
systems, recommender systems, social networks, and computer security.
\end{IEEEbiography}

\newpage

\begin{IEEEbiography}[{\includegraphics[width=1in,height=1.25in,clip,keepaspectratio]{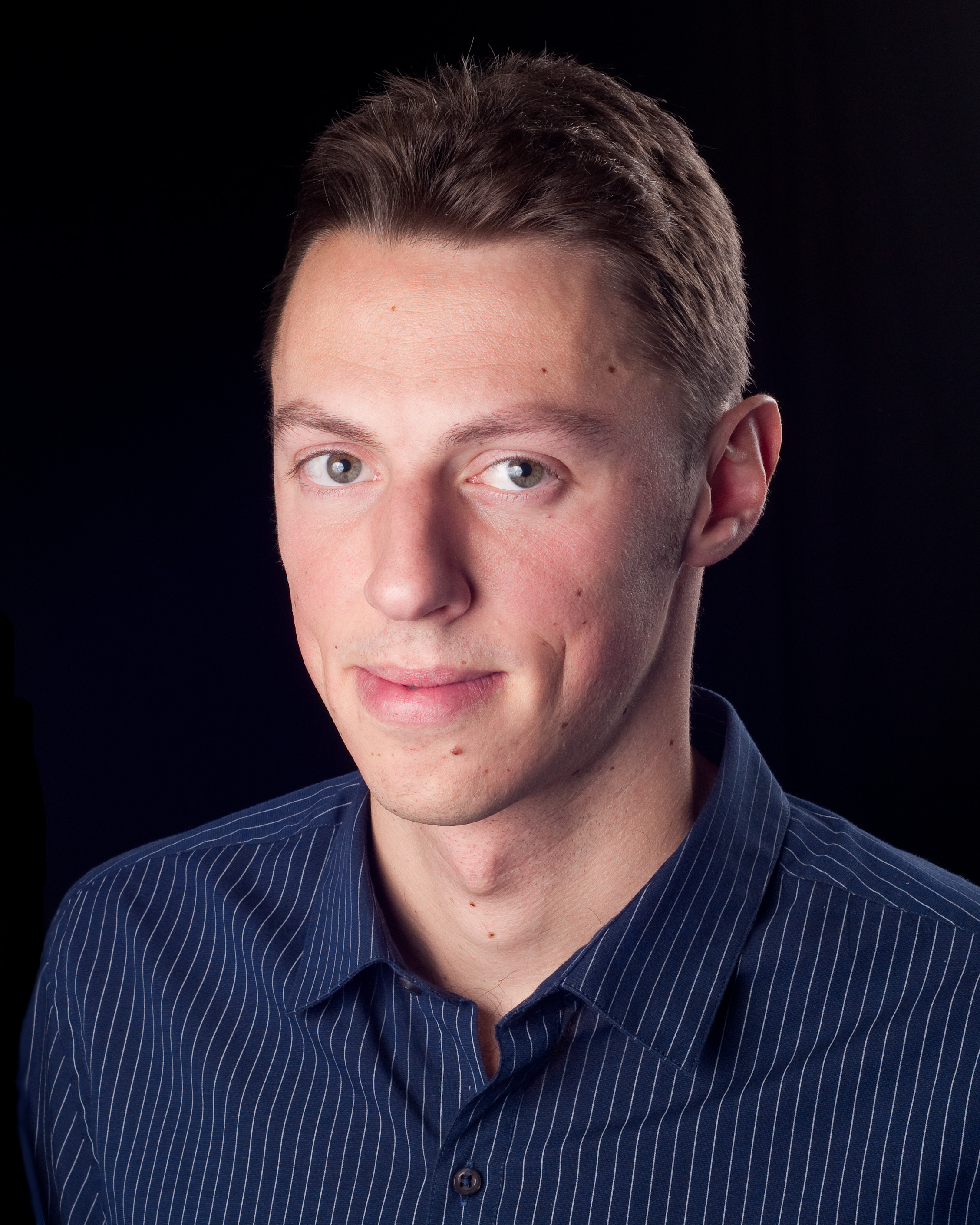}}]{Maurizio Filippone}
received a Master's degree in Physics and a
Ph.D. in Computer Science from the University of Genova, Italy, in
2004 and 2008, respectively.

In 2007, he was a Research Scholar with
George Mason University, Fairfax, VA.  From 2008 to 2011, he was a
Research Associate with the University of Sheffield, U.K. (2008-2009),
with the University of Glasgow, U.K. (2010), and with University
College London, U.K (2011).  He is currently a Lecturer with the
University of Glasgow, U.K.  His current research interests include
statistical methods for pattern recognition.

Dr Filippone serves as an Associate Editor for Pattern Recognition and
the IEEE Transactions on Neural Networks and Learning Systems.
\end{IEEEbiography}

\vfill

\begin{IEEEbiography}[{\includegraphics[width=1in,height=1.25in,clip,keepaspectratio]{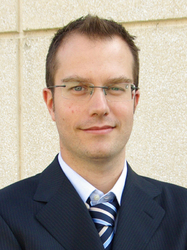}}]{Pietro Michiardi}
received his M.S. in Computer Science from EURECOM and his M.S. in
Electrical Engineering from Politecnico di Torino. Pietro received his
Ph.D. in Computer Science from Telecom ParisTech (former ENST, Paris),
and his HDR (Habilitation) from UNSA. Today, Pietro is an Assistant
professor of Computer Science at EURECOM, where he leads the
Distributed System Group, which blends theory and system research
focusing on large-scale distributed systems (including data processing
and data storage), and scalable algorithm design to mine massive
amounts of data. Additional research interests are on system,
algorithmic, and performance evaluation aspects of computer networks
and distributed systems.
\end{IEEEbiography}

\vfill

\begin{IEEEbiography}[{\includegraphics[width=1in,height=1.25in,clip,keepaspectratio]{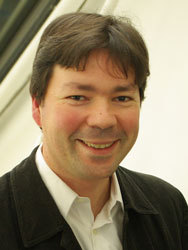}}]{Yves Roudier}
is an assistant professor in the Networking and Security
department. He received his PhD from the University of Nice Sophia
Antipolis in 1996. His expertise relates to network and system
security. He has developed an important expertise in various domains
of security, notably automotive security, Web Services security,
access control mechanisms, mobile code security, ubiquitous computing
security, and secure P2P storage. He is also investigating software
engineering for security. He has been investigating aspect-oriented
programming and binary code instrumentation to introduce security
mechanisms into distributed software and security requirements
specification. He is also interested in MDE-based approaches to
designing secure systems. He has authored more than 80 international
publications and was awarded 2 Best Paper awards.
\end{IEEEbiography}

\end{document}